\documentclass[a4paper,11pt]{article}
\usepackage{textcomp,amsmath,amsfonts,amssymb,amsthm,verbatim,mathrsfs}
\usepackage{graphicx}
\allowdisplaybreaks

\begin{document}

\title{Completeness of Hoare Logic over Nonstandard Models}

\author{
    Zhaowei Xu
    \footnotemark[1]
    \footnotemark[2],
    Yuefei Sui
    \footnotemark[3],
    Wenhui Zhang
    \footnotemark[1]
   }

\renewcommand{\thefootnote}{\fnsymbol{footnote}}

\footnotetext[1]{State Key Laboratory of Computer Science, Institute of Software, Chinese Academy of Sciences, Beijing, China}

\footnotetext[2]{Corresponding author at: University of Chinese Academy of Sciences, Beijing, China, xuzw@ios.ac.cn}

\footnotetext[3]{Key Laboratory of Intelligent Information Processing, Institute of Computing Technology, Chinese Academy of Sciences, Beijing, China}

\date{}

\maketitle

\begin{abstract}
 The nonstandard approach to program semantics has successfully resolved the completeness problem of Floyd-Hoare logic. The known versions of nonstandard semantics, the Hungary semantics and axiomatic semantics, are so general that they are absent either from mathematical elegance or from practical usefulness. The aim of this paper is to exhibit a not only mathematically elegant but also practically useful nonstandard semantics. A basic property of computable functions in the standard model $N$ of Peano arithmetic $PA$ is $\Sigma_1$-definability. However, the functions induced by the standard interpretation of while-programs $S$ in nonstandard models $M$ of $PA$ are not always arithmetical. The problem consists in that the standard termination of $S$ in $M$ uses the finiteness in $N$, which is not the finiteness in $M$. To this end, we shall give a new interpretation of $S$ in $M$ such that the termination of $S$ uses $M$-finiteness, and the functions produced by $S$ in all models of $PA$ have the uniform $\Sigma_1$-definability. Then we define, based on the new semantics of while-programs, a new semantics of Hoare logic in nonstandard models of $PA$, and show that the standard axiom system of Hoare logic is sound and complete w.r.t. the new semantics. It will be established, in $PA$, that the Hungary semantics and axiomatic semantics coincide with the new semantics of while-programs. Moreover, various comparisons with the previous results, usefulness of the nonstandard semantics, and remarks on the completeness issues are presented.

\smallskip

\end{abstract}

\section{Introduction}
Hoare logic is an axiomatic system of proving programs correct \cite{apt_1,floyd_1,hoare_1}, which has had a significant impact upon the methods of both designing and verifying programs \cite{mirkowska_1,harel_1,reynolds_1}. Hoare logic for the set $WP$ of while-programs with the first-order logical language $L$ and the first-order specification $T\subset L$ is denoted by $HL(T)$. Of special interest is the completeness of Hoare logic over nonstandard models.

The approach of using nonstandard models \cite{AvBBN14} in investigating Floyd-Hoare logic was started in \cite{andreka_1,andreka_2,andreka_3}, and pursued, refined by many authors since then, e.g., Csirmaz \cite{csirmaz_1}, Andreka et.al \cite{andreka_4}, Hortal\'{a}-Gonz\'{a}lez et.al \cite{hortala_1}, and Pasztor \cite{pasztor_1}. (For an extensive literature, see e.g. the references in \cite{makowsky_1}.) It's noteworthy that the Hungary semantics in \cite{andreka_3} is defined over a particular class of infinite sequences of vectors of numbers that should satisfy the induction axioms for all formulas, and Floyd logic is logically complete w.r.t. this semantics. In addition, according to the Floyd-Hoare principle \cite{floyd_1, hoare_1}, which states that the semantics of a programming language can be formally specified by the axioms and inference rules for proving the correctness of programs written in this language, an axiomatic semantics of while-programs has been suggested by Bergstra and Tucker \cite{bergstra_1} for which Hoare logic is logically complete. Observe that the Hungary semantics and axiomatic semantics are so general that they are absent either from mathematical elegance or from practical usefulness. It would be interesting, from a new perspective, to exhibit a not only mathematically elegant but also practically useful nonstandard semantics. In what follows, unless otherwise stated, let $L$ be the logical language of Peano arithmetic $PA$, and let $N$ be the standard model of $PA$.

The definability theorem \cite[Chapter 16]{c. and l.} says that a function in $N$ is computable (or equivalently $N$-computable \cite{rogers_1}) iff it is $\Sigma_1$-definable in $N$ (i.e., the graph relation of this function is $\Sigma_1$), which means that a computable function should be ``simple". The proof of this theorem only considers a finite computation in $N$, which can be encoded as a natural number in a uniform way. In nonstandard models $M$ of $PA$ \cite{model_1}, a while-program $S$ produces a (pseudo-$M$-computable) function $\vec{y} = f_S^M(\vec{x})$. Intuitively we hope that $f_S^M$ is simple in $M$, that is, $\Sigma_1$-definable in $M$. However, it is hopeless, because for the while-program $S ::= y:=0; while\ y<x\ do\ y:=y+1\ od$, $f_S^M$ is not $\Sigma_1$-definable, since the domain of $f_S^M$ is $N^M$, the standard part of $M$, which is not arithmetical by Robinson's overspill lemma \cite[Lemma 25.8]{c. and l.}. The problem consists in the interpretation of while-programs in $M$, where the termination uses the finiteness in $N$ (i.e. the usual finiteness), which is not the finiteness in $M$. We shall introduce the concept of $M$-finiteness for the finiteness in $M$ and show that every $M$-finite sequence of vectors of numbers of $M$ can be coded uniformly as a number of $M$. Then a new interpretation of while-programs $S$ in all models $M$ of $PA$ such that the termination of $S$ uses $M$-finiteness will be given (the new input-output relation of $S$ in $M$ is denoted by $A_S^M$ in order to distinguish from the standard one $R_S^M$). It will be shown that $A_S^M$ produces a simple function $\vec{y} = g_S^M(\vec{x})$ in $M$. On the other hand, there exists a $\Sigma_1$-formula $\alpha_S(\vec{x},\vec{y})\in L$, dependent on $S$ but independent of any model of $PA$, that defines an (abstract) function $\vec{y} = g_S(\vec{x})$ in $PA$, i.e., $PA\vdash \forall \vec{x},\vec{y},\vec{z} ( \alpha_S(\vec{x},\vec{y}) \wedge \alpha_S(\vec{x},\vec{z}/\vec{y}) \rightarrow \vec{y} = \vec{z} )$. (Actually the formula $\alpha_S(\vec{x},\vec{y})$ can be chosen as one defining the $N$-computable function $f_S^N$ produced by $S$ in $N$.) It will be found that the interpretation of $g_S$ in $M$ coincides with $g_S^M$, for every $M\models PA$. To illustrates this, the following commutative diagram is shown.
\begin{figure}[!ht]
  \centering
  \includegraphics[width = 1 in]{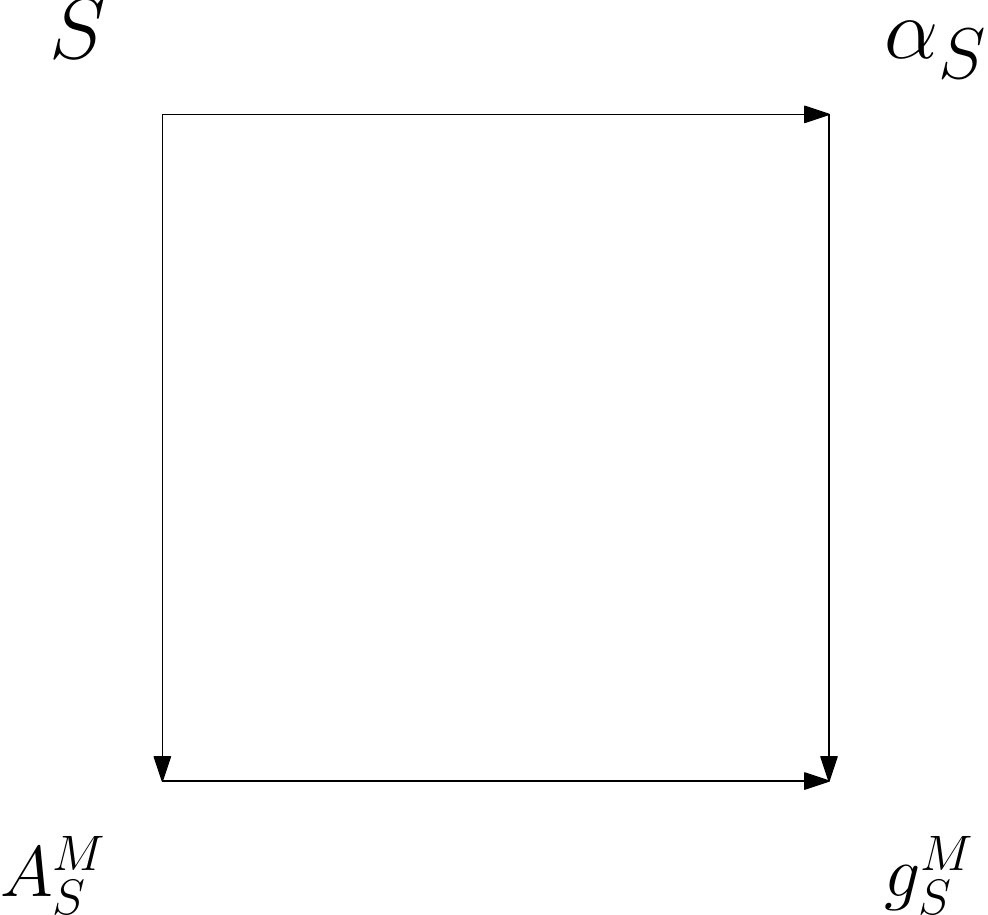}\\
  \caption{The commutative diagram for while-programs}\label{commutative_diagram}
\end{figure}
Thus, the new interpretation of while-programs extend $N$-computability from $N$ to nonstandard models of $PA$ with the uniform $\Sigma_1$-definability. We can now define $M$-computability to be the computability induced by the new semantics of while-programs in $M$. In Subsection \ref{comparison_with_the_standard_semantics}, the new semantics of while-programs is compared with the standard semantics; and Example \ref{example_for_two_semantics} shows that $(x,0),$ $(x,1),$ $\ldots,$ $(x,x)$ is the unique sequence of vectors of $M$ computable by $S ::= y:=0; while\ y<x\ do\ y:=y+1\ od$ with the input $(x,0)$ for the new semantics (i.e., $g_S^M(x,y) = (x,x)$).

We shall define, based on the new semantics of while-programs, a new semantics of Hoare logic in nonstandard models of $PA$. It will be proved that for any $PA\subseteq T \subseteq Th(N)$, the standard axiom system of $HL(T)$ is sound and complete w.r.t. the new semantics. It will be established, in $PA$, that the Hungary semantics and axiomatic semantics coincide with the new semantics of while-programs, and that the axiom system of Floyd logic is essentially the same as that of Hoare logic. Moreover, the reduction from $HL(T)$ to $T$ with $PA \subseteq T \subseteq Th(N)$ follows from the new completeness theorem of $HL(T)$, and the usefulness of the nonstandard semantics in Hoare logic is discussed. Finally, a detailed and thorough discussion of the completeness issues on Hoare logic is given.

The rest of this paper is structured as follows: some preliminary results are presented in Section 2; the concept of $M$-finiteness is introduced in Section 3; the definition of the new semantics of while-programs, and comparison with the standard and Hungary semantics are given in Section 4; the soundness and completeness of $HL(T)$ w.r.t. the new semantics, reduction from $HL(T)$ to $T$, and comparison with Floyd logic are shown in Section 5; the comparison with axiomatic semantics, and usefulness of nonstandard semantics in Hoare logic are shown in Section 6; remarks on the completeness issues of Hoare logic are given in Section 7; Section 8 concludes the paper.

\section{Preliminaries}

First some notations are introduced: in syntax, we write $\neg, \wedge, \vee, \rightarrow, \leftrightarrow, \forall, \exists$ to denote the negation, conjunction, disjunction, conditional, biconditional connectives and the universal, existential quantifiers; in semantics, we write $\sim, \&, |, \Rightarrow, \Leftrightarrow, \textbf{A}, \textbf{E}$ to denote the corresponding connectives and quantifiers.

\subsection{Peano arithmetic}

Let $L$ be the logical language of $PA$ with the signature $\Sigma = \{ +,\cdot,<,0,1 \}$. For simplicity, the sum of $1$ with itself $n$ times is abbreviated as $n$. We use $n$ to denote both a closed term and a natural number, and use $M$ to denote both a model and its domain, which will be clear from the context. The formula $\varphi(t/x)$ stands for the result of simultaneously substituting $t$ for free occurrences of $x$ in $\varphi$; and $\varphi(t/x)$ will be denoted $\varphi(t)$ if the default variable $x$ is obvious. The denotation of a term $t$ at an assignment $v$ (for all the first order variables) over a model $M$ of $L$, denoted $t^{M,v}$, receives the standard meaning. The satisfaction of a formula $\varphi\in L$ at an assignment $v$ over a model $M$ of $L$, denoted $M,v \models \varphi$, is defined as usual; the satisfaction of $\varphi$ in $M$, denoted $M\models \varphi$, is defined such that for any assignment $v$ over $M$, $M,v\models \varphi$; the satisfaction of a theory $T\subset L$ in $M$, denoted $M\models T$, is defined such that for any $\varphi\in T$, $M\models \varphi$; the satisfaction of $\varphi$ in a theory $T\subset L$, denoted $T\models \varphi$, is defined such that for any $M\models T$, $M\models \varphi$. And the derivation of a formula $\varphi\in L$ from a theory $T\subset L$, denoted $T\vdash \varphi$, is defined as usual. The set of all theorems of $T\subset L$, denoted $Thm(T)$, is defined to be $Thm(T) ::= \{\varphi \in L : T \vdash \varphi \}$. From Peano's axioms, one can deduce the induction axiom scheme $\varphi(0,\vec{y})\wedge \forall x\big(\varphi(x,\vec{y})\rightarrow \varphi(x+1,\vec{y})\big)\rightarrow \forall x\ \varphi(x,\vec{y})$, the bounded induction scheme $\varphi(0,\vec{y})\wedge \forall x<z \big(\varphi(x,\vec{y})\rightarrow \varphi(x+1,\vec{y})\big)\rightarrow \forall x\leq z\ \varphi(x,\vec{y})$ and the least-number principle $\exists x\ \varphi(x,\vec{y})\rightarrow \exists z ( \varphi(z,\vec{y})\wedge \forall u<z\ \neg \varphi(u,\vec{y}) )$, where $\varphi(x,\vec{y}) \in L$.

G\"{o}del's completeness theorem implies that for any $T\subset L$, $T\vdash \varphi$ iff $T\models \varphi$. The proof of this theorem shows that for any $T \subset L$ and any $\varphi\in L$, if $T\nvdash \varphi$ then there is a countable model $M$ of $L$ such that $M\models T$ and $M \nvDash \varphi$. This means that $T\models \varphi$ is equivalent to saying that for any countable $M\models T$, $M\models \varphi$ (the implication from left to right is trivial; for the opposite direction, assume that $T\nvDash \varphi$, by G\"{o}del's completeness theorem $T \nvdash \varphi$, then there exists a countable $M\models T$ such that $M\nvDash \varphi$, a contradiction), so in this paper only the countable models will be considered. Let $K$ be the set consisting of all natural numbers together with all pairs $(q,a)$ where $q$ is a rational number and $a$ an integer. Let $<_K$ be the order on $K$ in which the natural numbers come first, in their usual order, and the pairs afterward, ordered as follows: $(q,a) <_K (r,b)$ iff $q<r$ in the usual order on rational numbers, or $q=r$ and $a<b$ in the usual order on integers.
\newtheorem{order_relation_in_nonstandard_models}{Lemma}[subsection]
\begin{order_relation_in_nonstandard_models}[{Cf. \cite[p304]{c. and l.}}]\label{order_relation_in_nonstandard_models}
  The order relation on any countable nonstandard model of $PA$ is isomorphic to the ordering $<_K$ of $K$.
\end{order_relation_in_nonstandard_models}

By a $\Delta_0$-formula (or alternatively rudimentary formula) of $L$ we mean a formula built up from atomic formulas using only negation, conjunction, disjunction, and bounded quantifications $\forall x<t$ and $\exists x<t$, where $t$ is a term of $L$ (possibly involving free variables). (Conditionals and biconditionals are allowed, too, since these officially are just abbreviations for certain constructions involving negation, conjunction, and disjunction. So are the bounded quantifiers $\forall x\leq t$ and $\exists x\leq t$, since these are equivalent to $\forall x<t+1$ and $\exists x<t+1$.) By a $\Sigma_1$-formula (or alternatively $\exists$-rudimentary formula) we mean a formula of form $\exists x\ F$ where $F$ is a $\Delta_0$-formula. It holds, in $PA$, that every generalized $\Sigma_1$-formula, obtainable from $\Delta_0$-formulas by conjunction, disjunction, bounded quantifications, and unbounded existential quantification, is logically equivalent to a $\Sigma_1$-formula. (The equivalence in $N$ is shown in \cite[Proposition 16.10]{c. and l.}, and similarly for nonstandard models.)

\subsection{Coding functions}

\theoremstyle{definition}
\newtheorem{basic_functions}{Definition}[subsection]
\begin{basic_functions} \label{basic_functions}
The following definitions (cf. \cite[p54]{model_1}, \cite[p42]{computability_1}) are thought of as taking place in $PA$.\\
(a) $\Big \lfloor \frac{x}{y} \Big \rfloor$(the integer part of $x$ divided by $y$) is the two-place function
\begin{displaymath}
\Big\lfloor \frac{x}{y} \Big\rfloor = \left\{
                                \begin{array}{ll}
                                  \mbox{the unique $z$ s.t.} \\
                                  z\cdot y\leq x < (z+1)\cdot y, & \hbox{if $y\neq 0$;} \\
                                  0, & \hbox{if $y=0$.}
                                \end{array}
                              \right.
\end{displaymath}
(b) $\Big( \frac{x}{y} \Big)$(the remainder on dividing $x$ by $y$) is given by
\begin{displaymath}
 \Big( \frac{x}{y} \Big) = \left\{
                             \begin{array}{ll}
                               \mbox{the unique $z$ s.t.} \\
                               \exists w\leq x(y\cdot w+z=x\wedge z< y), & \hbox{if $y\neq 0$;} \\
                               x, & \hbox{otherwise.}
                             \end{array}
                           \right.
\end{displaymath}
(c) $x-y$ is defined by
\begin{displaymath}
x-y = \left\{
        \begin{array}{ll}
          \mbox{the unique $z$ s.t.\ } x=y+z, & \hbox{if $y\leq x$;} \\
          0, & \hbox{otherwise.}
        \end{array}
      \right.
\end{displaymath}
(d) $\sqrt{x}$ = the unique $y$ s.t. $y\cdot y \leq x < (y+1)\cdot (y+1)$.
\end{basic_functions}

The pairing function and unpairing functions in \cite[p45]{computability_1} build a one-to-one correspondence between the set of ordered pairs of natural numbers and the set of natural numbers, and similarly for nonstandard models of $PA$. We formulate this by the following.

\theoremstyle{definition}
\newtheorem{extended_basic_properties_of_extended_pairing_functions}[basic_functions]{Definition}
\begin{extended_basic_properties_of_extended_pairing_functions}
  Let $T_1(z) ::= \Big\lfloor \frac{\sqrt{8\cdot z+1}+1}{2}\Big\rfloor -1$ and let $T_2(z) ::= 2\cdot z-T_1(z)\cdot T_1(z)$. We define $\langle x,y\rangle$, $L(z)$ and $R(z)$ by
  \begin{eqnarray*}
    \langle x,y\rangle &=& \Big\lfloor \frac{(x+y+1)\cdot (x+y)}{2} \Big\rfloor + y, \\
    L(z) &=& \Big\lfloor\frac{T_2(z)-T_1(z)}{2}\Big\rfloor, \\
    R(z) &=& T_1(z)-L(z).
  \end{eqnarray*}
\end{extended_basic_properties_of_extended_pairing_functions}

\theoremstyle{plain}
\newtheorem{basic_properties_of_extended_pairing_functions}[basic_functions]{Lemma}
\begin{basic_properties_of_extended_pairing_functions}\label{basic_properties_of_extended_pairing_functions}
$PA$ proves that

(a) $\langle L(z),R(z)\rangle=z$;

(b) $L(\langle x,y\rangle)=x$;

(c) $R(\langle x,y\rangle)=y$.
\end{basic_properties_of_extended_pairing_functions}
Notice that $\langle x, y\rangle$ is called extended pairing function and $L(z)$, $R(z)$ extended unpairing functions, each of which extends the classical counterpart from $N$ to nonstandard models of $PA$. For notational convenience, $(L(z),R(z))$ is denoted by $\overline{z}$. The extended pairing and unpairing functions can be generalized to $n$-tuples (for each $n\in N$) by setting $\langle x_1,x_2,\ldots,x_n\rangle = \langle x_1,\langle x_2,\ldots,x_n\rangle\rangle$ and $\overline{\langle x_1,x_2,\ldots,x_n\rangle} = (x_1,\overline{\langle x_2,\ldots,x_n\rangle})$.

\theoremstyle{definition}
\newtheorem{extended_beta_function}[basic_functions]{Definition}
\begin{extended_beta_function}\label{extended_beta_function}
Define $(x)_i$ to be
\begin{equation*}
(x)_i=\Big( \frac{L(x)}{R(x)\cdot (i+ 1 )+1} \Big).
\end{equation*}
\end{extended_beta_function}

Such a function $(x)_i$ is usually called G$\ddot{o}$del's $\beta$-function, but in the present context we obviously prefer the more suggestive term extended $\beta$-function. It's easy to check, using Definition \ref{extended_beta_function}, that for every $M\models PA$ and every $w\in M$, there exists $i\in M$ such that $M\models \forall j> i ( (w)_j = (w)_i )$. By the Chinese Remainder Theorem, $(x)_i$ can be used to code every finite sequence of natural numbers by choosing a suitable (but not unique) value for $x$: for every finite sequence $a_0,a_1,\ldots,a_n$ of natural numbers, there exists a natural number $w$ such that $N\models (w)_i = a_i$ for all $i\leq n$ \cite[Theorem 2.4]{computability_1}. The following lemma formalizes this inside $PA$.

\theoremstyle{plain}
\newtheorem{basic_properties_of_extended_beta_function}[basic_functions]{Lemma}
\begin{basic_properties_of_extended_beta_function}[{Cf. \cite[p63]{model_1}}]\label{basic_properties_of_extended_beta_function}
$PA$ proves that

(a) $\forall x, \exists y\ (y)_0=x$;

(b) $\forall x,y,z, \exists w ( (\forall i<z\ (w)_i=(y)_i) \wedge (w)_z=x)$.
\end{basic_properties_of_extended_beta_function}

By Lemma \ref{basic_properties_of_extended_beta_function}, using mathematical induction, it follows that

\newtheorem{godel_lemma}[basic_functions]{Lemma}
\begin{godel_lemma}[G\"{o}del's Lemma, cf. {\cite[Lemma 5.9]{model_1}}]\label{godel_lemma}
Let $M\models PA$, and let $n\in N$ with $x_0$, $x_1$, $\ldots$, $x_{n-1}\in M$. Then there exists $w\in M$ such that $M \models (w)_i = x_i$ for all $i< n$.
\end{godel_lemma}

Note that these extended coding functions are all definable in $PA$ by $\Delta_0$-formulas of $L$.

\subsection{Hoare logic}

Based on the first-order logical language $L$, together with the program constructs ($:=$, $;$, $if$, $then$, $else$, $fi$, $while$, $do$, $od$), a while-program $S$ is defined by $S ::= x := E \mid S_1;S_2 \mid if\ B\ then\ S_1\ else\ S_2\ fi \mid while\ B\ do\ S_0\ od$, where an expression $E$ is defined by $E ::= 0 \mid 1 \mid x \mid E_1+E_2 \mid E_1\cdot E_2$, and a boolean expression $B$ is defined by $B ::= E_1 < E_2 \mid \neg B_1 \mid B_1\rightarrow B_2$. The set of all such while-programs is denoted $WP$. Without further declaration, for a while-program $S\in WP$, the vector $(x_1, x_2, \ldots, x_n)$ of all $n$ program variables $x_1,$ $x_2,$ $\ldots,$ $x_n$ occurring in $S$ will be denoted by $\vec{x}$; the vector $(a_1,a_2,\ldots,a_n)$ of $n$ numbers $a_1,$ $a_2,$ $\ldots,$ $a_n\in M$ will be denoted by $\vec{a}$; the connectives will be assumed to distribute over the components of the vectors (for instance, $\vec{a}\in M$ means $a_1,$ $a_2,$ $\ldots,$ $a_n\in M$, $\vec{x} = \vec{y}$ means $\bigwedge_{i = 1}^{n} x_i = y_i$, and $\varphi(\vec{t}/\vec{x})$ means $\varphi(t_1/x_1,t_2/x_2,\ldots,t_n/x_n)$). For an assignment $v$ over a model $M$ of $L$, let $v(\vec{x})$ be the vector of elements of $M$ assigned to $\vec{x}$ at $v$, and let $v(\vec{y}/\vec{x})$ be an assignment as $v$ except that $v(\vec{y}/\vec{x})(\vec{x}) = \vec{y}$. For every $S\in WP$ and every model $M$ of $L$, the standard input-output relation $R_S^M$ of $S$ in $M$ is a binary relation on the set of assignments over $M$ inductively defined as follows:
\begin{itemize}
  \item $(v,v')\in R_{x:=E}^{M}$ $\Leftrightarrow$ $v'=v(E^{M,v}/x)$;
  \item $(v,v')\in R_{S_1;S_2}^{M}$ $\Leftrightarrow$ $(v,v')\in R_{S_1}^{M}\circ R_{S_2}^{M}$, where $(z,z')\in R_1\circ R_2$ $\Leftrightarrow$ $\textbf{E} z'' ( (z,z'')\in R_1$ $\&$ $(z'',z')\in R_2 )$;
  \item $(v,v')\in R_{if\ B\ then\ S_1\ else\ S_2\ fi}^{M}$ $\Leftrightarrow$ $M,v\models B$ $\&$ $(v,v')\in R_{S_1}^{M}$ $|$ $M,v\not\models B$ $\&$ $(v,v')\in R_{S_2}^{M}$;
  \item $(v,v')\in R_{while\ B\ do\ S_{0}\ od}^{M}$ $\Leftrightarrow$ $\textbf{E} i\in N$, $\textbf{E}\vec{x_0},\ldots,\vec{x_i} \in M$ $( v(\vec{x}) = \vec{x_0}$ $\&$ $\textbf{A} j < i ( M,v(\vec{x_j}/\vec{x})\models B$ $\&$ $(v(\vec{x_j}/\vec{x}), v(\vec{x_{j+1}}/\vec{x}))\in R_{S_0}^{M} ) $ $\&$ $v' = v(\vec{x_i}/\vec{x})$ $\&$ $M,v'\not\models B )$.
\end{itemize}
Given $S\in WP$ and a model $M$ of $L$, $R_S^M$ produces a vectorial function $\vec{y} = f_S^{M}(\vec{x})$ such that for any $\vec{a},\vec{b}\in M$, $f_S^M(\vec{a}) = \vec{b} \Leftrightarrow \textbf{E} v,v' ( v(\vec{x})=\vec{a}\ \&\ v'(\vec{x})=\vec{b}\ \&\ (v,v')\in R_S^M )$. Given a model $M$ of $L$ and an asserted program $\{p\}S\{q\}$, $\{p\}S\{q\}$ is satisfied at $M$ w.r.t. the standard semantics, denoted $M\models_S \{p\}S\{q\}$, iff for any assignment $v$ over $M$, $M,v\models p \Rightarrow \textbf{A}v'\big((v,v')\in R_S^M \Rightarrow M,v'\models q\big)$. Given a theory $T\subset L$ and an asserted program $\{p\}S\{q\}$, $\{p\}S\{q\}$ is satisfied at $T$ w.r.t. the standard semantics, denoted by $HL(T)\models_S \{p\}S\{q\}$, iff for any $M\models T$, $M \models_S \{p\}S\{q\}$.

Hoare logic $HL(T)$ for the set $WP$ of while-programs with the first-order logical language $L$ and the first-order specification $T\subset L$ has the standard axioms and inference rules \cite[Section 2]{apt_1}: let $p, q, p_1, q_1, r \in L$, and let $S,$ $S_0,$ $S_1,$ $S_2 \in WP$.

(1) Assignment Axiom Scheme: for $E$ an expression of $L$ and $x$ a variable of $L$, the asserted program
\begin{displaymath}
\begin{array}{l}
\{p(E/x)\}x:=E\{p\}
\end{array}
\end{displaymath}
is an axiom.

(2) Composition Rule:
\begin{displaymath}
\begin{array}{c}
  \{p\}S_1\{r\},\{r\}S_2\{q\} \\
  \hline
  \{p\}S_1;S_2\{q\}
\end{array}
\end{displaymath}

(3) Conditional Rule:
\begin{displaymath}
\begin{array}{c}
  \{p\wedge B\}S_1\{q\},\{p\wedge \neg B\}S_2\{q\} \\
  \hline
  \{p\}if\ B\ then\ S_1\ else\ S_2\ fi\{q\}
\end{array}
\end{displaymath}

(4) Iteration Rule:
\begin{displaymath}
\begin{array}{c}
  \{r\wedge B\}S_0\{r\} \\
  \hline
  \{r\}while\ B\ do\ S_0\ od\{r\wedge \neg B\}
\end{array}
\end{displaymath}

(5) Consequence Rule:
\begin{displaymath}
\begin{array}{c}
  p\rightarrow p_1,\{p_1\}S\{q_1\},q_1\rightarrow q \\
  \hline
  \{p\}S\{q\}
\end{array}
\end{displaymath}

And, in connection with (5),

(6) Specification Axiom: each theorem of $T$ is an axiom.

We write $HL(T)\vdash \{p\}S\{q\}$ to denote that the asserted program $\{p\}S\{q\}$ is provable from $HL(T)$. By the logical completeness of $HL(T)$ w.r.t. the standard semantics is meant that for any asserted program $\{p\}S\{q\}$, $HL(T)\vdash \{p\}S\{q\}$ iff $HL(T)\models_S \{p\}S\{q\}$.

\section{The concept of $M$-finiteness}

To begin with, the notion of pseudo-$M$-finiteness is introduced.

\theoremstyle{definition}
\newtheorem{definition_of_pseudo_M_finiteness}{Definition}[section]
\begin{definition_of_pseudo_M_finiteness}
  Let $M\models PA$. A sequence of numbers of $M$ is pseudo-$M$-finite if it is indexed by an initial segment of $M$. That is, every sequence of numbers $a_0,a_1,\ldots,a_i\in M$ with $i\in M$ is pseudo-$M$-finite.
\end{definition_of_pseudo_M_finiteness}

The sequences as defined above behave like finite sequences as far as the underlying model is concerned, but they may be infinite sequences when viewed from the outside world: let $i\in M\models PA$ be nonstandard; the pseudo-$M$-finite sequence of numbers $0$, $1$, $\ldots$, $i$ is infinite. In particular, every pseudo-$N$-finite (or equivalently finite) sequence of natural numbers can be coded as a natural number by extended $\beta$-function. However, this property doesn't hold for nonstandard models.

\theoremstyle{plain}
\newtheorem{pseudo_$M$_finite_is_uncodified}[definition_of_pseudo_M_finiteness]{Proposition}
\begin{pseudo_$M$_finite_is_uncodified}\label{pseudo_$M$_finite_is_uncodified}
  Every countable nonstandard model $M$ of $PA$ has a pseudo-$M$-finite sequence of numbers that can't be coded as a number by extended $\beta$-function.
\end{pseudo_$M$_finite_is_uncodified}
\begin{proof}
  Let $i\in M$ be nonstandard. We consider the set $A$ $=$ $\{$ $<$ $a_0$, $a_1$, $\ldots$, $a_i$ $>$ $:$ for each $j\leq i$, $a_j = 0$ or $1$ $\}$ of pseudo-$M$-finite sequences of numbers of $M$. It's obvious that $A$ is uncountable. However, $M$ has only a countable number of elements used as codes. This means that one of elements of $A$ can't be coded as a number of $M$ by extended $\beta$-function.
\end{proof}

\theoremstyle{definition}
\newtheorem{definition_of_M_finiteness}[definition_of_pseudo_M_finiteness]{Definition}
\begin{definition_of_M_finiteness}\label{definition_of_M_finiteness}
  Let $M\models PA$. A pseudo-$M$-finite sequence $\vec{a_0}$, $\vec{a_1}$, $\ldots$, $\vec{a_i}$ of vectors of numbers of $M$ is $M$-finite if there exist $\psi(\vec{x},y,\vec{z}) \in L$ (the arguments $\vec{x}$, $y$ could be dummy) and $\vec{a}\in M$ such that for all $j$ smaller than or equal to $i$ there is a unique $\vec{z}\in M$ satisfying $M\models \psi(\vec{a},j,\vec{z})$, and this $\vec{z}$ happens to be $\vec{a_j}$, in symbols $M\models \forall j\leq i ( \exists! \vec{z}\ \psi(\vec{a},j,\vec{z}) \wedge \psi(\vec{a},j,\vec{a_j}) )$.
\end{definition_of_M_finiteness}

For the sequence $\vec{x_0}$, $\vec{x_1}$, $\ldots$, $\vec{x_n}\in M\models PA$ with $n\in N$, by Lemmas \ref{basic_properties_of_extended_pairing_functions} and \ref{godel_lemma}, there exist $\psi(x,y,\vec{z}) ::= \overline{(x)_y} =  \vec{z}$ and $x\in M$ such that $M\models \forall j\leq n ( \exists! \vec{z}\ \psi(x,j,\vec{z}) \wedge \psi(x,j,\vec{x_j}) )$.  For the sequence $0$, $1$, $\ldots$, $i$ with $i\in M\models PA$ being nonstandard, there exist $\psi(x,y,z) ::= z = y$ and $x\in M$ such that $M\models \forall j\leq i ( \exists! z\ \psi(x,j,z) \wedge \psi(x,j,j) )$. Observe from the above that the concept of finiteness is equal to that of $N$-finiteness (in symbols, finiteness $=$ $N$-finiteness) and, for any nonstandard $M\models PA$, the concept of finiteness is strictly less than that of $M$-finiteness (in symbols, finiteness $<$ $M$-finiteness).

The following theorem shows that for any $M\models PA$, every $M$-finite sequence of vectors of numbers of $M$ can be coded as a number of $M$ by extended coding functions.

\theoremstyle{plain}
\newtheorem{code_generated_by_formulas}[definition_of_pseudo_M_finiteness]{Theorem}
\begin{code_generated_by_formulas}\label{code_generated_by_formulas}
Let $\psi(\vec{x},y,\vec{z})\in L$ and $M \models PA$. Then we have
\begin{eqnarray*}
  M &\models& \forall \vec{x},y ( \forall i\leq y, \exists!\vec{z}\ \psi(\vec{x},i,\vec{z}) \\
   &&  \rightarrow \exists w, \forall i\leq y, \forall \vec{z} ( \psi(\vec{x},i,\vec{z}) \rightarrow \overline{(w)_i} =  \vec{z} ) ).
\end{eqnarray*}
\end{code_generated_by_formulas}
\begin{proof}
Fix $\vec{x}\in M$. It suffices to prove that $M \models \forall y ( \forall i\leq y, \exists!\vec{z}\ \psi(\vec{x},i,\vec{z}) \rightarrow \exists w, \forall i\leq y, \forall \vec{z} ( \psi(\vec{x},i,\vec{z}) \rightarrow \overline{(w)_i} =  \vec{z} ) )$ by induction on $y$ in the formula $\forall i\leq y, \exists!\vec{z}\ \psi(\vec{x},i,\vec{z}) \rightarrow \exists w, \forall i\leq y, \forall \vec{z} ( \psi(\vec{x},i,\vec{z}) \rightarrow \overline{(w)_i} = \vec{z} )$.

Suppose that $M\models \forall i\leq 0, \exists!\vec{z}\ \psi(\vec{x},i,\vec{z})$ for $y=0$. Then there exists a unique $\vec{z}\in M$ such that $M\models \psi(\vec{x},0,\vec{z})$. Using Lemmas \ref{basic_properties_of_extended_pairing_functions} and \ref{basic_properties_of_extended_beta_function}, there is $w\in M$ such that $M\models \overline{(w)_0} = \vec{z}$. It follows that $M\models \exists w, \forall i\leq 0, \forall \vec{z} ( \psi(\vec{x},i,\vec{z}) \rightarrow \overline{(w)_i} = \vec{z} )$. Thus we have
\begin{eqnarray*}
  M &\models& \forall i\leq 0, \exists!\vec{z}\ \psi(\vec{x},i,\vec{z}) \\
   &&  \rightarrow \exists w, \forall i\leq 0, \forall \vec{z} ( \psi(\vec{x},i,\vec{z}) \rightarrow \overline{(w)_i} = \vec{z} ).
\end{eqnarray*}

As the induction hypothesis, suppose
\begin{eqnarray*}
  M &\models& \forall i\leq y, \exists!\vec{z}\ \psi(\vec{x},i,\vec{z}) \\
   &&  \rightarrow \exists w, \forall i\leq y, \forall \vec{z} ( \psi(\vec{x},i,\vec{z}) \rightarrow \overline{(w)_i} = \vec{z} )
\end{eqnarray*}
for any $y\in M$. Then we consider the case for $y + 1$. Assume that $M\models \forall i\leq y+1, \exists!\vec{z}\ \psi(\vec{x},i,\vec{z})$. That is, there exists a unique sequence $\vec{z_0},\ldots,\vec{z_y},\vec{z_{y+1}}\in M$ such that $M\models \forall i\leq y+1\ \psi(\vec{x},i,\vec{z_i})$. By the induction hypothesis there exists $w\in M$ such that $M\models \forall i\leq y, \forall \vec{z} ( \psi(\vec{x},i,\vec{z}) \rightarrow \overline{(w)_i} = \vec{z} )$. That is, $M\models \forall i\leq y\ \overline{(w)_i} = \vec{z_i}$. Using Lemmas \ref{basic_properties_of_extended_pairing_functions} and \ref{basic_properties_of_extended_beta_function}, there exists $w'\in M$ such that $M\models \forall i<y+1\ (w')_i = (w)_i \wedge \overline{(w')_{y+1}} = \vec{z_{y+1}}$ and hence we have $M\models \exists w, \forall i\leq y+1, \forall \vec{z} ( \psi(\vec{x},i,\vec{z}) \rightarrow \overline{(w)_i} = \vec{z} )$, so it follows that
\begin{eqnarray*}
   M&\models& \forall i\leq y+1, \exists!\vec{z}\ \psi(\vec{x},i,\vec{z}) \\
   && \rightarrow \exists w, \forall i\leq y+1, \forall \vec{z} ( \psi(\vec{x},i,\vec{z}) \rightarrow \overline{(w)_i} = \vec{z} ).
\end{eqnarray*}
\end{proof}

It's noteworthy, in Definition \ref{definition_of_M_finiteness}, that the uniqueness condition $\exists! \vec{z}$ in the assertion $M\models \forall j\leq i ( \exists! \vec{z}\ \psi(\vec{a},j,\vec{z}) \wedge \psi(\vec{a},j,\vec{a_j}) )$ is extremely important: let $\psi(\vec{x},y,z) ::= (z = 0) \vee (z = 1)$; each element $<a_0,a_1,\ldots,a_i>$ of $A$, as defined in the proof of Proposition \ref{pseudo_$M$_finite_is_uncodified}, satisfies the property $M\models \forall j\leq i\ \psi(\vec{a},j,a_j)$ for arbitrary $\vec{a}\in M$; yet there exists such an element that can't be coded as a number of $M$.

Defining $\psi(x,y,\vec{z})$ by $\psi(x,y,\vec{z}) ::= \overline{(x)_y} =  \vec{z}$, it's easy to see that every pseudo-$M$-finite sequence of vectors of numbers of $M$ coded as a number of $M$ by extended coding functions is $M$-finite. Hence the concept of $M$-finiteness captures precisely what can be coded in $M$ by extended coding functions; so it could be taken as an ideal extension to the usual concept of finiteness. We can say, in the sense of the above, that Theorem \ref{code_generated_by_formulas} extends G\"{o}del's Lemma (cf. Lemma \ref{godel_lemma}) into a general setting.

\section{The X-semantics of while-programs}
\subsection{The definition of the X-semantics}

\theoremstyle{definition}
\newtheorem{X_semantics_of_while_programs}{Definition}[subsection]
\begin{X_semantics_of_while_programs}[The X-semantics of while-programs]\label{X_semantics_of_while_programs}
Let $M\models PA$, and $S\in WP$ with program variables $\vec{x} = (x_1,x_2,\ldots,x_n)$. The new input-output relation $A_S^M$ for $S$ on the set of assignments over $M$, and the formula $\alpha_S(\vec{x},\vec{y})\in L$ for $S$, where $\vec{y} = (y_1,y_2,\ldots,y_n)$ is disjoint from $\vec{x}$, are defined inductively in (a) and (b) as follows:

Assignment: $S \equiv x_i := E$.

(a) $(v,v')\in A_S^M$ $\Leftrightarrow$ $v'$ = $v(E^{M,v}/x_i)$.

(b) $\alpha_S(\vec{x},\vec{y}) ::= y_i=E(\vec{x}) \wedge \bigwedge_{1\leq j \leq n}^{j \neq i} y_j=x_j$.

Composition: $S\equiv S_1;S_2$.

(a) $(v,v')\in A_S^M$ $\Leftrightarrow$ $(v,v')\in A_{S_1}^M\circ A_{S_2}^M$.

(b) $\alpha_S(\vec{x},\vec{y}) ::= \exists \vec{z} (\alpha_{S_1}(\vec{x},\vec{z}/\vec{y})\wedge \alpha_{S_2}(\vec{z}/\vec{x},\vec{y}))$.

Conditional: $S\equiv if\ B\ then\ S_1\ else\ S_2\ fi$.

(a) $(v,v')\in A_S^M$ $\Leftrightarrow$ $M,v\models B$ $\&$ $(v,v')\in A_{S_1}^M$ $|$ $M,v\not\models B$ $\&$ $(v,v')\in A_{S_2}^M$.

(b) $\alpha_S(\vec{x},\vec{y}) ::= (B(\vec{x})\wedge \alpha_{S_1}(\vec{x},\vec{y})) \vee (\neg B(\vec{x}) \wedge \alpha_{S_2}(\vec{x},\vec{y}))$.

Iteration: $S\equiv while\ B\ do\ S_0\ od$.

We define the formula $\varphi_{S_0}(\vec{x},y,\vec{z})\in L$ by
\begin{equation*}
\varphi_{S_0}(\vec{x},y,\vec{z}) ::= \exists w ( \overline{(w)_0} = \vec{x} \wedge \forall j<y\ \alpha_{S_0}(\overline{(w)_j},\overline{(w)_{j+1}}) \wedge \overline{(w)_y} = \vec{z} ).
\end{equation*}

(a) $(v,v')\in A_S^M$ $\Leftrightarrow$ $\textbf{E}$ $i,$ $\vec{x_0},$ $\vec{x_1},$ $\ldots,$ $\vec{x_i}\in M$ $[$ $M\models \forall j\leq i\ \varphi_{S_0}(v(\vec{x}),j,\vec{x_j})$ $\&$ $v(\vec{x}) = \vec{x_0}$ $\&$ $\textbf{A}$ $j< i$ $($ $M,v(\vec{x_j}/\vec{x})\models B$ $\&$ $(v(\vec{x_j}/\vec{x}), v(\vec{x_{j+1}}/\vec{x}) ) \in A_{S_0}^M$ $)$ $\&$ $v'(\vec{x}) = \vec{x_i}$ $\&$ $M,v'\not\models B$ $]$.

(b) We first let
\begin{eqnarray*}
  C_S(i,w,\vec{x},\vec{y}) &::=& \vec{x}=\overline{(w)_0} \wedge \forall j<i (B(\overline{(w)_j}/\vec{x}) \\
                         &&       \wedge\: \alpha_{S_0}(\overline{(w)_j}/\vec{x},\overline{(w)_{j+1}}/\vec{y})) \wedge \vec{y} = \overline{(w)_i}
\end{eqnarray*}
then set
\begin{equation*}
\alpha_S^*(i,\vec{x},\vec{y}) ::= \exists w\ C_S(i,w,\vec{x},\vec{y})
\end{equation*}
and finally define
\begin{equation*}
\alpha_S(\vec{x},\vec{y}) ::= \exists i\ \alpha_S^*(i,\vec{x},\vec{y}) \wedge \neg B(\vec{y}/\vec{x}).
\end{equation*}
\end{X_semantics_of_while_programs}

Note that, in Definition \ref{X_semantics_of_while_programs}, the definitions of $A_S^M$ and $\alpha_S(\vec{x},\vec{y})$ proceed simultaneously by induction on $S$, in that for the iteration case, $\varphi_{S_0}(\vec{x},y,\vec{z})$ is constructed from $\alpha_{S_0}$, and the definition of $A_S^M$ refers to $\varphi_{S_0}(\vec{x},y,\vec{z})$; for the role of $\varphi_{S_0}(\vec{x},y,\vec{z})$, we shall show in Claim \ref{computable_sequence_generated_by_formulas} that any sequence $\vec{x_0},$ $\vec{x_1},$ $\ldots,$ $\vec{x_i}$ of vectors of numbers of $M$ such that $M\models \forall j\leq i\ \varphi_{S_0}(v(\vec{x}),j,\vec{x_j})$ is $M$-finite. Note also that a particular $M$-finite sequence of vectors of numbers of $M$ is adopted here rather than the more natural one, that is the sequence $\vec{x_0},$ $\vec{x_1},$ $\ldots,$ $\vec{x_i}$ such that $M \models \vec{x_0} = \vec{x} \wedge \forall j<i ~ \alpha_{S_0}(\vec{x_j},\vec{x_{j+1}})$ with $i,\vec{x}\in M$, because it has not been proved or disproved that the latter can be coded as a number of $M$ by extended coding functions. (If such a sequence satisfies the induction axioms for all formulas, then it is $M$-finite; for more details, cf. Definition \ref{definition_of_M_continuous_sequences} and Theorem \ref{M_continuous_computation_sequence_is_codified}.)

\theoremstyle{plain}
\newtheorem{involved_formulas_are_sigma_1}[X_semantics_of_while_programs]{Proposition}
\begin{involved_formulas_are_sigma_1}
  For every $S\in WP$, it is the case that in $PA$, $\alpha_S(\vec{x},\vec{y})$ and $\varphi_S(\vec{x},y,\vec{z})$ are logically equivalent to a $\Sigma_1$-formula.
\end{involved_formulas_are_sigma_1}
\begin{proof}
  Let $S\in WP$. It's easy to check, by induction on $S$, that $\alpha_S(\vec{x},\vec{y})$ and $\varphi_S(\vec{x},y,\vec{z})$ are generalized $\Sigma_1$-formulas, so, in $PA$, each is logically equivalent to a $\Sigma_1$-formula.
\end{proof}

\theoremstyle{definition}
\newtheorem{M_computable_functions}[X_semantics_of_while_programs]{Definition}
\begin{M_computable_functions}\label{M_computable_functions}
For every $S\in WP$ and every $M\models PA$, $\vec{y} = g_S^M(\vec{x})$ is defined such that for any $\vec{a},\vec{b}\in M$, $g_S^M(\vec{a}) = \vec{b} \Leftrightarrow \textbf{E} v,v' ( v(\vec{x})=\vec{a}\ \&\ v'(\vec{x})=\vec{b}\ \&\ (v,v')\in A_S^M )$.
\end{M_computable_functions}

\theoremstyle{plain}
\newtheorem{arithmetical_definability_of_M_computable_functions}[X_semantics_of_while_programs]{Theorem}
\begin{arithmetical_definability_of_M_computable_functions}\label{arithmetical_definability_of_M_computable_functions}
For every $S\in WP$ and every $M\models PA$, $\vec{y} = g_S^M(\vec{x})$ is a function such that for any $\vec{a},\vec{b}\in M$, $g_S^M(\vec{a}) = \vec{b}$ iff $M \models \alpha_S(\vec{a},\vec{b})$.
\end{arithmetical_definability_of_M_computable_functions}

\begin{proof}
To prove the theorem, by Definition \ref{M_computable_functions}, it suffices to prove that, for every $S\in WP$ and every $M\models PA$, $(v,v')\in A_S^M$ iff $M \models \alpha_S(v(\vec{x}),v'(\vec{x}))$ in (a), and $M \models \forall \vec{x},\vec{y},\vec{z}(\alpha_S(\vec{x},\vec{y})\wedge\alpha_S(\vec{x},\vec{z}/\vec{y})\rightarrow \vec{y}=\vec{z})$ in (b). Fix $M\models PA$. The argument proceeds by induction on $S$.

Assignment: $S \equiv x_i := E$.

(a) Consider $(v,v')\in A_S^M$ as follows: by the definition of $A_S^M$, it is equivalent to $v'(x_i) = E^{M,v}$ $\&$ $\textbf{A}$ $1 \leq j\leq n$, $j\neq i$ $v'(x_j) = v(x_j)$; then it is equivalent to $M \models v'(x_i) = E(v(\vec{x})/\vec{x})$ $\&$ $\textbf{A}$ $1\leq i \leq n$, $j\neq i$ $v'(x_j) = v(x_j)$; by the definition of $\alpha_S(\vec{x},\vec{y})$, it is equivalent to $M \models \alpha_S(v(\vec{x}),v'(\vec{x}))$.

(b) It's trivial that $M \models \forall \vec{x},\vec{y},\vec{z}(\alpha_S(\vec{x},\vec{y})\wedge\alpha_S(\vec{x},\vec{z}/\vec{y})\rightarrow \vec{y}=\vec{z})$.

Composition: $S\equiv S_1;S_2$.

(a) Consider $(v,v')\in A_S^M$ as follows: by the definition of $A_S^M$, it is equivalent to $(v,v')\in A_{S_1}^M\circ A_{S_2}^M$; by the definition of $\circ$, it is equivalent to $\textbf{E}v''$ $[$ $(v,v'')\in A_{S_1}^M$ $\&$ $(v'',v')\in A_{S_2}^M$ $]$; by the induction hypothesis, it is equivalent to $\textbf{E}$ $v''$ $[$ $M \models \alpha_{S_1}(v(\vec{x}),v''(\vec{x}))$ $\&$ $M \models \alpha_{S_2}(v''(\vec{x}),v'(\vec{x}))$ $]$; by the definition of $\alpha_S(\vec{x},\vec{y})$, it is equivalent to $M \models \alpha_S(v(\vec{x}),v'(\vec{x}))$.

(b) Applying the induction hypothesis to $S_1$ (resp. $S_2$) yields $M \models \forall \vec{x},\vec{y},\vec{z}(\alpha_{S_1}(\vec{x},\vec{y})\wedge\alpha_{S_1}(\vec{x},\vec{z}/\vec{y})\rightarrow \vec{y}=\vec{z})$ (resp. $M \models \forall \vec{x},\vec{y},\vec{z}(\alpha_{S_2}(\vec{x},\vec{y})\wedge\alpha_{S_2}(\vec{x},\vec{z}/\vec{y})\rightarrow \vec{y}=\vec{z})$). By the definition of $\alpha_S(\vec{x},\vec{y})$, we have that $M \models \forall \vec{x},\vec{y},\vec{z}(\alpha_S(\vec{x},\vec{y})\wedge\alpha_S(\vec{x},\vec{z}/\vec{y})\rightarrow \vec{y}=\vec{z})$.

Conditional: $S\equiv if\ B\ then\ S_1\ else\ S_2\ fi$.

(a) Consider $(v,v')\in A_S^M$ as follows: by the definition of $A_S^M$, it is equivalent to $M,v\models B$ $\&$ $(v,v')\in A_{S_1}^M$ $|$ $M,v\not\models B$ $\&$ $(v,v')\in A_{S_2}^M$; by the induction hypothesis, it is equivalent to $M\models B(v(\vec{x}))$ $\&$ $M \models \alpha_{S_1}(v(\vec{x}),v'(\vec{x}))$ $|$ $M\not\models B(v(\vec{x}))$ $\&$ $M \models \alpha_{S_2}(v(\vec{x}),v'(\vec{x}))$; by the definition of $\alpha_S(\vec{x},\vec{y})$, it is equivalent to $M \models \alpha_S(v(\vec{x}),v'(\vec{x}))$.

(b) Similar to the composition case.

Iteration: $S\equiv while\ B\ do\ S_0\ od$.

We begin with

\newtheorem{computable_sequence_generated_by_formulas}[X_semantics_of_while_programs]{Claim}
\begin{computable_sequence_generated_by_formulas}\label{computable_sequence_generated_by_formulas}
Let $i, \vec{x}, \vec{x_0},$ $\vec{x_1},$ $\ldots,$ $\vec{x_i} \in M$ with $M\models \forall j \leq i\ \varphi_{S_0}(\vec{x},j,\vec{x_j})$. Then it is the case that
  \begin{description}
    \item[(1)] $M\models \forall j \leq i ( \exists! \vec{z}\ \varphi_{S_0}(\vec{x},j,\vec{z}) \wedge \varphi_{S_0}(\vec{x},j,\vec{x_j}) )$;
    \item[(2)] $M\models \exists w, \forall j \leq i\ \overline{(w)_j} = \vec{x_j}$.
  \end{description}
\end{computable_sequence_generated_by_formulas}

The proof of Claim \ref{computable_sequence_generated_by_formulas} is shown as follows:

Proof of (1). By the premise, it suffices to prove the uniqueness of $\vec{z}$ in $M\models \forall j \leq i, \exists \vec{z}\ \varphi_{S_0}(\vec{x},j,\vec{z})$. suppose
\begin{eqnarray*}
  j\leq i,w,w'\in M &\models& \overline{(w)_0} = \overline{(w')_0} = \vec{x} \wedge \forall k<j\ \alpha_{S_0}(\overline{(w)_k},\overline{(w)_{k+1}}) \\
   && \wedge \: \forall k<j\ \alpha_{S_0}(\overline{(w')_k},\overline{(w')_{k+1}}),
\end{eqnarray*}
by the induction hypothesis we have $M \models \forall \vec{x},\vec{y},\vec{z}(\alpha_{S_0}(\vec{x},\vec{y})\wedge\alpha_{S_0}(\vec{x},\vec{z}/\vec{y})\rightarrow \vec{y}=\vec{z})$, then by induction on $k$ up to $j$ we can see that $M\models \forall k\leq j\ \overline{(w)_k} = \overline{(w')_k}$ and hence in particular $M\models \overline{(w)_j} = \overline{(w')_j}$, so it follows that $M\models \forall j \leq i, \vec{z},\vec{z'} ( \varphi_{S_0}(\vec{x},j,\vec{z}) \wedge \varphi_{S_0}(\vec{x},j,\vec{z'}) \rightarrow \vec{z} = \vec{z'} )$.

Proof of (2). Immediate from (1) and Theorem \ref{code_generated_by_formulas}.

We next give the main proof of this case:

(a) Consider $(v,v')\in A_S^M$ as follows: by the definition of $A_S^M$, it is equivalent to $\textbf{E}$ $i,$ $\vec{x_0},$ $\vec{x_1},$ $\ldots,$ $\vec{x_i}\in M$ $[$ $M\models \forall j\leq i ~ \varphi_{S_0}(v(\vec{x}),j,\vec{x_j})$ $\&$ $v(\vec{x}) = \vec{x_0}$ $\&$ $\textbf{A}$ $j< i$ $($ $M,v(\vec{x_j}/\vec{x})\models B$ $\&$ $(v(\vec{x_j}/\vec{x}), v(\vec{x_{j+1}}/\vec{x}) ) \in A_{S_0}^M$ $)$ $\&$ $v'(\vec{x}) = \vec{x_i}$ $\&$ $M,v'\not\models B$ $]$; by the induction hypothesis, it is equivalent to $\textbf{E}$ $i,$ $\vec{x_0},$ $\vec{x_1},$ $\ldots,$ $\vec{x_i}\in M$ $[$ $M\models \forall j\leq i ~ \varphi_{S_0}(v(\vec{x}),j,\vec{x_j})$ $\&$ $v(\vec{x}) = \vec{x_0}$ $\&$ $M \models \forall j< i(B(\vec{x_j})\wedge \alpha_{S_0}(\vec{x_j},\vec{x_{j+1}}) )$ $\&$ $v'(\vec{x}) = \vec{x_i}$ $\&$ $M \not\models B(v'(\vec{x}))$ $]$; using Claim \ref{computable_sequence_generated_by_formulas}, it is equivalent to $\textbf{E}$ $i,w\in M$ $[$ $M\models v(\vec{x}) = \overline{(w)_0}$ $\&$ $M\models \forall j<i ( B(\overline{(w)_j}) \wedge \alpha_{S_0}(\overline{(w)_j}, \overline{(w)_{j+1}}) )$ $\&$ $M\models v'(\vec{x}) = \overline{(w)_i}$ $\&$ $M\not\models B(v'(\vec{x}))$ $]$; by the definition of $\alpha_S(\vec{x},\vec{y})$, it is equivalent to $M\models \alpha_S(v(\vec{x}),v'(\vec{x}))$.

(b) Suppose $\vec{a}, \vec{b}, \vec{c}\in M$ with $M \models\alpha_S(\vec{a},\vec{b}) \wedge \alpha_S(\vec{a},\vec{c})$. Then we have to prove that $\vec{b} = \vec{c}$. By the supposition, together with the definition of $\alpha_S$, there exist $i,w,i',w'\in M$ such that
\begin{align}
  M \models\ &\vec{a}=\overline{(w)_0} \wedge \forall j<i (B(\overline{(w)_j}) \nonumber \\
            &\wedge\: \alpha_{S_0}(\overline{(w)_j},\overline{(w)_{j+1}})) \wedge \vec{b} = \overline{(w)_i} \wedge \neg B(\vec{b}) \tag{1}
\end{align}
and
\begin{align}
  M \models\ & \vec{a}=\overline{(w')_0} \wedge \forall j<i' (B(\overline{(w')_j}) \nonumber \\
             & \wedge\: \alpha_{S_0}(\overline{(w')_j},\overline{(w')_{j+1}})) \wedge \vec{c} = \overline{(w')_{i'}} \wedge \neg B(\vec{c}). \tag{2}
\end{align}
To prove $\vec{b} = \vec{c}$, by (1) and (2), it is enough to prove that $i = i'$ and $M \models \overline{(w)_i}=\overline{(w')_{i'}}$. By the order relation of $M$ (cf. Lemma \ref{order_relation_in_nonstandard_models}), it follows that $i\leq^{M} i'$ or $i' \leq^{M} i$. Without loss of generality, let $i\leq^{M} i'$. By (1) and (2), we have that $M \models \overline{(w)_0}=\overline{(w')_0}$. Suppose $j <^{M} i$ and $M \models \overline{(w)_j}=\overline{(w')_j}$, then by the induction hypothesis we immediately have $M \models \forall \vec{x}, \vec{y},\vec{z}(\alpha_{S_0}(\vec{x},\vec{y})\wedge\alpha_{S_0}(\vec{x},\vec{z}/\vec{y}) \rightarrow \vec{y}=\vec{z})$ and hence by (1), (2) it follows that $M \models \overline{(w)_{j+1}}=\overline{(w')_{j+1}}$, so we can see $M \models \forall j \leq i\ \overline{(w)_j}=\overline{(w')_j}$ by induction on $j$ up to $i$. Suppose for a contradiction that $i <^{M} i'$. This, together with (2), implies $M \models B(\overline{(w')_i})$. From (1), we immediately have $M \not\models B(\overline{(w)_i})$. This, together with the fact $M \models \overline{(w)_i}=\overline{(w')_i}$, implies $M \not\models B(\overline{(w')_i})$, a contradiction. Thus we have $i = i'$ and finally it follows that $M \models \overline{(w)_i}=\overline{(w')_{i'}}$.
\end{proof}

Note that in the proof of Theorem \ref{arithmetical_definability_of_M_computable_functions}, the proof for that $(v,v')\in A_S^M$ iff $M \models \alpha_S(v(\vec{x}),v'(\vec{x}))$, and the proof for that $M \models \forall \vec{x},\vec{y},\vec{z}(\alpha_S(\vec{x},\vec{y})\wedge\alpha_S(\vec{x},\vec{z}/\vec{y})\rightarrow \vec{y}=\vec{z})$ proceed simultaneously by induction on $S$, in that for the iteration case, the proof for that $(v,v')\in A_S^M$ iff $M \models \alpha_S(v(\vec{x}),v'(\vec{x}))$ employs Claim \ref{computable_sequence_generated_by_formulas}, whose proof uses the induction hypothesis $M \models \forall \vec{x},\vec{y},\vec{z}(\alpha_{S_0}(\vec{x},\vec{y})\wedge\alpha_{S_0}(\vec{x},\vec{z}/\vec{y})\rightarrow \vec{y}=\vec{z})$.

Let $S\equiv while\ B\ do\ S_0\ od$ and $M\models PA$. The X-semantics of while-programs (cf. Definition \ref{X_semantics_of_while_programs}) shows that every run $\vec{x_0},$ $\vec{x_1},$ $\ldots,$ $\vec{x_i}$ of $S$ in $M$ should satisfy the constraint $M\models \forall j\leq i\ \varphi_{S_0}(\vec{x_0},j,\vec{x_j})$. By Claim \ref{computable_sequence_generated_by_formulas}, on giving the input $\vec{x_0}$, this run will be uniquely determined by the constraint. Thus the natural way to decide whether a given sequence $\vec{x_0},$ $\vec{x_1},$ $\ldots,$ $\vec{x_i}\in M$ (typically given by a formula with the argument $\vec{x_0}$) is a run of $S$ is to prove $M\models \forall j\leq i\ \varphi_{S_0}(\vec{x_0},j,\vec{x_j})$ by induction on $j$ up to $i$ in the formula $\varphi_{S_0}(\vec{x_0},j,\vec{x_j})$ using Lemmas \ref{basic_properties_of_extended_pairing_functions} and \ref{basic_properties_of_extended_beta_function}.

\newtheorem{code_recursively_generated}[X_semantics_of_while_programs]{Proposition}
\begin{code_recursively_generated}\label{code_recursively_generated_label}
Let $\psi(\vec{x},\vec{y}) \in L$ and $M\models PA$ with $M \models \forall \vec{x},\vec{y},\vec{z} (\psi(\vec{x},\vec{y}) \wedge \psi(\vec{x},\vec{z})\rightarrow \vec{y}=\vec{z})$. Then for any $\vec{x}\in M$, we have that either
\begin{equation}
  M \models \forall i, \exists w ( \overline{(w)_0} = \vec{x} \wedge \forall j<i\ \psi(\overline{(w)_j},\overline{(w)_{j+1}}) ) \tag{3}
\end{equation}
or
\begin{equation}
  M \models \exists i,w( \overline{(w)_0} = \vec{x} \wedge \forall j<i\ \psi(\overline{(w)_j},\overline{(w)_{j+1}}) \wedge \neg \exists \vec{z}\ \psi(\overline{(w)_i}, \vec{z})). \tag{4}
\end{equation}
\end{code_recursively_generated}
\begin{proof}
Fix $\vec{x}\in M$. Suppose that (3) doesn't hold. Then we have to prove (4). By the supposition, we have that $M \models \exists i, \neg\exists w ( \overline{(w)_0} = \vec{x} \wedge \forall j<i\ \psi(\overline{(w)_j},\overline{(w)_{j+1}}) )$.  Using the least number principle (cf. the Preliminaries), there exists $i\in M$ such that
\begin{equation}
  M \models \neg \exists w ( \overline{(w)_0} = \vec{x} \wedge \forall j<i\ \psi(\overline{(w)_j},\overline{(w)_{j+1}}) ) \tag{5}
\end{equation}
and
\begin{equation}
  M \models \forall k<i, \exists w ( \overline{(w)_0} = \vec{x} \wedge \forall j<k\ \psi(\overline{(w)_j},\overline{(w)_{j+1}}) ). \tag{6}
\end{equation}
Then $M\models i > 0$, otherwise contradicting (5) using Lemma \ref{basic_properties_of_extended_beta_function}, and hence by (6) there exists $w\in M$ such that $M \models \overline{(w)_0} = \vec{x} \wedge \forall j<i-1\ \psi(\overline{(w)_j},\overline{(w)_{j+1}})$. To prove (4), it suffices to prove $M\models \neg\exists \vec{z}\ \psi(\overline{(w)_{i-1}}, \vec{z})$. Assume for a contradiction that $M\nvDash \neg\exists \vec{z}\ \psi(\overline{(w)_{i-1}}, \vec{z})$. Then there exists $\vec{z}\in M$ such that $M\models \psi(\overline{(w)_{i-1}}, \vec{z})$, using Lemmas \ref{basic_properties_of_extended_pairing_functions} and \ref{basic_properties_of_extended_beta_function} there exists $w'\in M$ such that $M\models \forall j<i\ (w')_j=(w)_j \wedge \overline{(w')_i} = \vec{z}$, hence $M \models \overline{(w')_0} = \vec{x} \wedge \forall j<i\ \psi(\overline{(w')_j},\overline{(w')_{j+1}})$, a contradiction to (5), and so (4) holds.
\end{proof}

\newtheorem{computational_power_for_the_nonstandard_semantics}[X_semantics_of_while_programs]{Corollary}
\begin{computational_power_for_the_nonstandard_semantics}\label{computational_power_for_the_nonstandard_semantics}
Let $M\models PA$, $v$ be an assignment over $M$, and $S ::= while\ B\ do\ S_0\ od$ with $S_0\in WP$. Then either of the following two cases holds.
\begin{description}
  \item[(1)] $\textbf{A}$ $i\in M$, $\textbf{E}$ $\vec{x_0},$ $\vec{x_1},$ $\ldots,$ $\vec{x_i}\in M$ $[$ $M\models \forall j\leq i ~ \varphi_{S_0}(v(\vec{x}),j,\vec{x_j})$ $\&$ $v(\vec{x}) = \vec{x_0}$ $\&$ $\textbf{A}$ $j< i$ $(v(\vec{x_j}/\vec{x}), v(\vec{x_{j+1}}/\vec{x}) ) \in A_{S_0}^M$ $]$;
  \item[(2)] $\textbf{E}$ $i,$ $\vec{x_0},$ $\vec{x_1},$ $\ldots,$ $\vec{x_i}\in M$ $[$ $M\models \forall j\leq i ~ \varphi_{S_0}(v(\vec{x}),j,\vec{x_j})$ $\&$ $v(\vec{x}) = \vec{x_0}$ $\&$ $\textbf{A}$ $j< i$ $(v(\vec{x_j}/\vec{x}), v(\vec{x_{j+1}}/\vec{x}) )\in A_{S_0}^M$ $\&$ $\sim \textbf{E}$ $\vec{z}\in M$ $( v(\vec{x_i}/\vec{x}), v(\vec{z}/\vec{x}) ) \in A_{S_0}^M$ $]$.
\end{description}
\end{computational_power_for_the_nonstandard_semantics}
\begin{proof}
By Proposition \ref{code_recursively_generated_label}, the definition of $\varphi_{S_0}$ (cf. Definition \ref{X_semantics_of_while_programs}), and the fact that for any $S\in WP$ and any $M\models PA$, $(v,v')\in A_S^M$ iff $M \models \alpha_S(v(\vec{x}),v'(\vec{x}))$, and $M \models \forall \vec{x},\vec{y},\vec{z}(\alpha_S(\vec{x},\vec{y})\wedge\alpha_S(\vec{x},\vec{z}/\vec{y})\rightarrow \vec{y}=\vec{z})$ (cf. Theorem \ref{arithmetical_definability_of_M_computable_functions}).
\end{proof}

Observe from Corollary \ref{computational_power_for_the_nonstandard_semantics} that in Definition \ref{X_semantics_of_while_programs}, for the iteration case, the definition of $A_S^M$, though restricted, can still exhaust the sequences of vectors of numbers of $M$ computable by $S_0$ that we are concerned with.

\newtheorem{induction_of_m_finiteness}[X_semantics_of_while_programs]{Proposition}
\begin{induction_of_m_finiteness}[The induction of $M$-finiteness]\label{induction_of_m_finiteness}
Let $\vec{x},\vec{y},\vec{z}\in M \models PA$ and $S_0\in WP$. Then it is the case that
\begin{description}
  \item[(i)] $M \models \varphi_{S_0}(\vec{x},0,\vec{y}) \leftrightarrow \vec{x} = \vec{y}$;
  \item[(ii)] $M \models \varphi_{S_0}(\vec{x},i,\vec{y}) \wedge \alpha_{S_0}(\vec{y},\vec{z}) \leftrightarrow \varphi_{S_0}(\vec{x},i+1,\vec{z})$.
\end{description}
\end{induction_of_m_finiteness}
\begin{proof}
  (i) Consider $M \models \varphi_{S_0}(\vec{x},0,\vec{y})$ as follows: by the definition of $\varphi_{S_0}$, it is equivalent to saying that there exists $w\in M$ such that $M \models \overline{(w)_0} = \vec{x} \wedge \overline{(w)_0} = \vec{y}$; using Lemmas \ref{basic_properties_of_extended_pairing_functions} and \ref{basic_properties_of_extended_beta_function}, it is equivalent to $M\models \vec{x}=\vec{y}$.

  (ii) Consider $M \models \varphi_{S_0}(\vec{x},i,\vec{y}) \wedge \alpha_{S_0}(\vec{y},\vec{z})$ as follows. By the definition of $\varphi_{S_0}$, it is equivalent to saying that there exists $w\in M$ such that $M \models \overline{(w)_0} = \vec{x} \wedge \forall j<i\ \alpha_{S_0}(\overline{(w)_j},\overline{(w)_{j+1}}) \wedge \overline{(w)_i} = \vec{y} \wedge \alpha_{S_0}(\vec{y},\vec{z})$. Using Lemmas \ref{basic_properties_of_extended_pairing_functions} and \ref{basic_properties_of_extended_beta_function}, it is equivalent to saying that there exist $w, w'\in M$ such that $M \models \overline{(w)_0} = \vec{x} \wedge \forall j<i\ \alpha_{S_0}(\overline{(w)_j},\overline{(w)_{j+1}}) \wedge \overline{(w)_i} = \vec{y} \wedge \alpha_{S_0}(\vec{y},\vec{z})$ and $M \models \forall j<i+1\ (w')_j=(w)_j \wedge \overline{(w')_{i+1}} = \vec{z}$. Using Claim \ref{computable_sequence_generated_by_formulas} ($\Leftarrow$. Let $w=w'$. $\overline{(w)_i}$ is the unique value of $\vec{y}$), it is equivalent to saying that there exists $w'\in M$ such that $M \models \overline{(w')_0} = \vec{x} \wedge \forall j<i+1\ \alpha_{S_0}(\overline{(w')_j},\overline{(w')_{j+1}}) \wedge \overline{(w')_{i+1}} = \vec{z}$. By the definition of $\varphi_{S_0}$, it is equivalent to $M \models \varphi_{S_0}(\vec{x},i+1,\vec{z})$.
\end{proof}

\newtheorem{induction_of_loop_programs}[X_semantics_of_while_programs]{Corollary}
\begin{induction_of_loop_programs}[The induction of loop programs]\label{induction_of_loop_programs}
Let $\vec{x},\vec{y},\vec{z}\in M\models PA$, and $S$ $::=$ $while$ $B$ $do$ $S_0$ $od$ with $S_0\in WP$. Then it is the case that

(i) $M\models \alpha_S^*(0,\vec{x},\vec{y}) \leftrightarrow \vec{x}=\vec{y}$;

(ii) $M\models \alpha_S^*(i,\vec{x},\vec{y})\wedge ( B(\vec{y}/\vec{x})\wedge \alpha_{S_0}(\vec{y}/\vec{x},\vec{z}/\vec{y}) ) \leftrightarrow \alpha_S^*(i+1,\vec{x},\vec{z}/\vec{y})$.
\end{induction_of_loop_programs}
\begin{proof}
By Proposition \ref{induction_of_m_finiteness}, together with the definitions of $\varphi_{S_0}$ and $\alpha_S^*$ (cf. Definition \ref{X_semantics_of_while_programs}).
\end{proof}

\subsection{Comparison with the standard semantics}\label{comparison_with_the_standard_semantics}

In this subsection, the connections between the standard semantics and X-semantics of while-programs are built: to illustrate the similarity and difference of the two semantics, an example is shown; for the general relationship of the two semantics, two propositions are given ; the reasons why the standard semantics is ``complex'' in logic are discussed.

\theoremstyle{definition}
\newtheorem{example_for_two_semantics}{Example}[subsection]
\begin{example_for_two_semantics}[Identity functions]\label{example_for_two_semantics}
  Let $S ::= y:=0; while\ y<x\ do\ y:=y+1\ od$. For every $x,y\in N$, $f_S^N(x,y)$ = $g_S^N(x,y)$ = $(x,x)$. For every nonstandard $M\models PA$, and every $x,y\in M$, $f_S^M(x,y) = (x,x)$ if $x$ is a standard number of $M$, and $f_S^M$ is undefined for $(x,y)$ otherwise; in contrast, $g_S^M(x,y) = (x,x)$.
\end{example_for_two_semantics}
\begin{proof}
  Fix $x,y \in M\models PA$. The value of $f_S^M$ for the input $(x,y)$ is easy to calculate. It remains to show that $g_S^M(x,y) = (x,x)$. Let $S_0 ::= y:=y+1$. To prove $g_S^M(x,y) = (x,x)$, by X-semantics of while-programs (cf. Definition \ref{X_semantics_of_while_programs}), it suffices to prove that $M\models \forall y\leq x\ \varphi_{S_0}((x, 0), y, (x, y))$ by induction on $y$ up to $x$ in the formula $\varphi_{S_0}((x, 0), y, (x, y))$. Using Lemmas \ref{basic_properties_of_extended_pairing_functions} and \ref{basic_properties_of_extended_beta_function}, we have $M\models \exists w\ \overline{(w)_0} = ( x,0 )$, so it follows that $M\models \varphi_{S_0}((x, 0), 0, (x, 0))$. Suppose $x > y\in M$ and $M\models \varphi_{S_0}((x, 0), y, (x, y))$. By definition of $\varphi_{S_0}$, there exists $w\in M$ such that $M\models \overline{(w)_0} = (x, 0) \wedge \forall j<y\ \alpha_{S_0}(\overline{(w)_j},\overline{(w)_{j+1}}) \wedge \overline{(w)_y} = ( x,y )$. Using Lemmas \ref{basic_properties_of_extended_pairing_functions} and \ref{basic_properties_of_extended_beta_function}, there is $w'\in M$ such that $M\models  \forall j<y+1\ (w')_j = (w)_j \wedge \overline{(w')_{y+1}} = ( x, y+1 )$, so $M\models \varphi_{S_0}((x, 0), y+1, (x, y+1))$ follows.
\end{proof}

\theoremstyle{plain}
\newtheorem{comparison_of_two_semantics_over_N}[example_for_two_semantics]{Proposition}
\begin{comparison_of_two_semantics_over_N}\label{comparison_of_two_semantics_over_N}
For every $S\in WP$, $R_S^N = A_S^N$.
\end{comparison_of_two_semantics_over_N}
\begin{proof}
Let $S\in WP$. To prove $R_S^N = A_S^N$, by Theorem \ref{arithmetical_definability_of_M_computable_functions}, it suffices to prove that $(v,v')\in R_S^N$ iff $N \models \alpha_S(v(\vec{x}), v'(\vec{x}))$, which is easy for the reader to verify.
\end{proof}

\newtheorem{comparison_of_two_semantics_over_M}[example_for_two_semantics]{Proposition}
\begin{comparison_of_two_semantics_over_M}\label{comparison_of_two_semantics_over_M}
  Let $M\models PA$ be nonstandard. For every $S\in WP$, $R_S^M \subseteq A_S^M$. Moreover, for particular choices of $S$, both $R_S^M = A_S^M$ and $R_S^M \subsetneqq A_S^M$ are able to hold.
\end{comparison_of_two_semantics_over_M}
\begin{proof}
  Let $S\in WP$. The proof for $R_S^M \subseteq A_S^M$ proceeds by induction on $S$: the cases for assignment, composition and conditional are easy; for the iteration case, by Theorem \ref{arithmetical_definability_of_M_computable_functions}, it suffices to prove that $(v,v')\in R_S^M$ $\Rightarrow$ $M \models \alpha_S(v(\vec{x}), v'(\vec{x}))$; this is the case due to Corollary \ref{induction_of_loop_programs}. Moreover, for $S \equiv x := x + 1$, we have that $R_S^M = A_S^M$; for $S$ as defined in Example \ref{example_for_two_semantics}, we can see that $R_S^M \subsetneqq A_S^M$.
\end{proof}

Perhaps the most significant difference between the two semantics is that the X-semantics in nonstandard models is ``simple'' (or equivalently $\Sigma_1$-definable) while the standard semantics in nonstandard models is not always so. It appears that Robinson's Overspill Lemma, which says that the finiteness in $N$ (defined by $\textbf{E} i \in N$) is not arithmetical, leads to this. Actually, the finiteness in $N$ is not defined in first-order logic: introducing more first-order ingredients into $PA$ is of no help.

\newtheorem{standard_numbers_is_not_definable_in_first_order_logic}[example_for_two_semantics]{Theorem}
\begin{standard_numbers_is_not_definable_in_first_order_logic}[Undefinability of the finiteness in $N$]\label{standard_numbers_is_not_definable_in_first_order_logic}
  The set of standard numbers is not defined in first-order logic.
\end{standard_numbers_is_not_definable_in_first_order_logic}
\begin{proof}
Suppose that $T$ is any first-order consistent extension of $PA$. Then we have to prove that the set of standard numbers is not defined in $T$. Assume for a contradiction that there exists $\theta(x)$ in the language of $T$ that defines the set of standard numbers in all models of $T$. As in the existence proof of nonstandard models of arithmetic, add a constant $\infty$ to the language of $T$ and apply the Compactness Theorem \cite[Theorem 12.15]{c. and l.} to the theory
  \begin{equation*}
    T\cup \{ \theta(\infty) \} \cup \{ \infty \neq n : n\in N \}
  \end{equation*}
  to conclude that it has a model. The denotation of $\infty$ in any such model will be a nonstandard element, contradicting the assumption.
\end{proof}

\subsection{Comparison with the Hungary semantics}\label{comparison_with_the_Hungary_semantics}

By a labeled program we understand a finite sequence of commands of the type $(i:u)$ ($i\in N$ is the label, and $u$ is either an assignment statement $x\hookleftarrow E$ or a conditional goto statement $if\ B\ goto\ j$, where $E$, $B$ are defined as for while-programs, and $j\in N$), in which no two members have the same label. Formally, a labeled program $P$ can be written as $P$ $=$ $< (i_0, u_0), (i_1, u_1), \ldots, (i_n, u_n) >$, where $\textbf{A} a < b \leq n\ i_a \neq i_b$. Let $M\models PA$. A trace of $P$ in $M$ is a sequence $s ::= < \vec{s_a} >_{a\in M}$ indexed by the elements of $M$ such that $\vec{s_a}$ is a vector of numbers of $M$ assigned to the program variables of $P$ (including the control variable $\lambda$ used to denote the labels), and $s$ is a run of $P$; $P$ is said to halt at $\vec{s_i}$ if $\vec{s_i}(\lambda)\not\in \{i_m\ :\ m\leq n\}$ and $\textbf{A}j \geq^M i\ \vec{s_j} = \vec{s_i}$. (For more details, the reader refers to \cite{andreka_3}.)

\theoremstyle{definition}
\newtheorem{definition_of_continuous_sequences}{Definition}[subsection]
\begin{definition_of_continuous_sequences}[Cf. {\cite[pp165]{andreka_3}}]\label{definition_of_continuous_sequences}
  Let $M\models PA$. The sequence $< \vec{s_a} >_{a\in M}$ in $M$ is continuous if it satisfies the induction axiom scheme, that is, for any $\varphi(\vec{x})\in L$,
  \begin{equation*}
    M \models \varphi(\vec{s_0}) \wedge \forall x ( \varphi(\vec{s_x}) \rightarrow \varphi(\vec{s_{x+1}}) ) \rightarrow \forall x\ \varphi(\vec{s_x}).
  \end{equation*}
\end{definition_of_continuous_sequences}

By a continuous trace of $P$ in $M$ we understand a trace $< \vec{s_a} >_{a\in M}$ of $P$ which is continuous. The semantics of labeled programs over continuous traces is called the Hungary semantics.

Note that a continuous sequence in $M$ is $M$-infinite, which means that it is indexed by all elements of $M$, so it might not be coded as a number by extended coding functions (consider, e.g., the sequence $< a >_{a\in M}$). As while-programs are structured (by contrast labeled programs are unstructured), this definition of continuous sequences (cf. Definition \ref{definition_of_continuous_sequences}) is not appropriate for building the continuous semantics of while-programs. In the present context, we prefer to adopt the following definition of continuous sequences.

\newtheorem{definition_of_M_continuous_sequences}[definition_of_continuous_sequences]{Definition}
\begin{definition_of_M_continuous_sequences}\label{definition_of_M_continuous_sequences}
  Let $i\in M\models PA$. A pseudo-$M$-finite sequence $\vec{a_0}$, $\vec{a_1}$, $\ldots$, $\vec{a_i}$ of vectors of numbers of $M$ is $M$-continuous if it satisfies the induction axiom scheme, that is, for any $\varphi(\vec{x},y,\vec{z})\in L$ and any $\vec{x}\in M$,
  \begin{equation*}
    M \models \varphi(\vec{x},0,\vec{a_0}) \wedge \forall j < i ( \varphi(\vec{x},j,\vec{a_j}) \rightarrow \varphi(\vec{x},j+1,\vec{a_{j+1}}) ) \rightarrow \forall j \leq i\ \varphi(\vec{x},j,\vec{a_j}).
  \end{equation*}
\end{definition_of_M_continuous_sequences}

It is inessential that the parameters $\vec{x}$, $y$ in $\varphi(\vec{x},y,\vec{z})$ are explicit in Definition \ref{definition_of_M_continuous_sequences}, yet not in Definition \ref{definition_of_continuous_sequences}. In particular, every finite sequence of vectors of natural numbers is $N$-continuous.

\theoremstyle{plain}
\newtheorem{M_continuous_is_uncodified}[definition_of_continuous_sequences]{Proposition}
\begin{M_continuous_is_uncodified}\label{M_continuous_is_uncodified}
  Every countable nonstandard model $M$ of $PA$ has an $M$-continuous sequence of numbers that cannot be coded as a number by extended $\beta$-function.
\end{M_continuous_is_uncodified}
\begin{proof}
  Let $i\in M$ be nonstandard. Consider the set $A$ $=$ $\{$ $<$ $a_0$, $a_1$, $\ldots$, $a_i$ $>$ : $a_0 = 0$, $a_1 = 1$, and for each $2 \leq j\leq i$, $a_j = 0$ or $1$ $\}$. It's easy to check that each element of $A$ is $M$-continuous. Note that $A$ is uncountable. However, $M$ has only countably many elements used as codes. This means that one of elements of $A$ can't be coded as a number of $M$ by extended $\beta$-function.
\end{proof}

\newtheorem{M_finiteness_is_M_continuous}[definition_of_continuous_sequences]{Proposition}
\begin{M_finiteness_is_M_continuous}\label{M_finiteness_is_M_continuous}
  Let $M\models PA$. Every $M$-finite sequence $\vec{a_0}$, $\vec{a_1}$, $\ldots$, $\vec{a_i}$ of vectors of numbers of $M$ is $M$-continuous.
\end{M_finiteness_is_M_continuous}
\begin{proof}
  Follows from the definition of $M$-finiteness (cf. Definition \ref{definition_of_M_finiteness}) and the induction axiom scheme of $PA$.
\end{proof}

Observe from Theorem \ref{code_generated_by_formulas} and Propositions \ref{M_continuous_is_uncodified}, \ref{M_finiteness_is_M_continuous} that the concept of $M$-finiteness is strictly less than that of $M$-continuousness (in symbols, $M$-finiteness $<$ $M$-continuousness).

In order to distinguish from the known semantics of while-programs, for the continuous semantics of while-programs, we use $C_S^M$ to denote the input-output relation of $S\in WP$ in $M\models PA$. For any $S\in WP$ and any $M\models PA$, $C_S^M$ is defined as in the standard semantics except for the iteration case:

$(v,v')\in C_{while\ B\ do\ S_0\ od}^M$ $\Leftrightarrow$ $\textbf{E}$ an $M$-continuous sequence $\vec{x_0}$, $\ldots$, $\vec{x_i} \in M$ $[$ $v(\vec{x}) = \vec{x_0}$ $\&$ $\textbf{A} j < i ( M,v(\vec{x_j}/\vec{x})\models B$ $\&$ $(v(\vec{x_j}/\vec{x}), v(\vec{x_{j+1}}/\vec{x}))\in C_{S_0}^M ) $ $\&$ $v' = v(\vec{x_i}/\vec{x})$ $\&$ $M,v'\not\models B$ $]$.

It's not difficult to prove the equivalence of the Hungary semantics to the continuous semantics of while-programs: the equivalence in $N$ is shown in \cite[Section 3-3.1]{manna_1}; and similarly for nonstandard models. To show the equivalence of the Hungary semantics to the X-semantics of while-programs, it suffices to show the equivalence of the continuous semantics to the X-semantics.

\newtheorem{M_continuous_computation_sequence_is_codified}[definition_of_continuous_sequences]{Theorem}
\begin{M_continuous_computation_sequence_is_codified}\label{M_continuous_computation_sequence_is_codified}
  Let $M\models PA$ and $\psi(\vec{x},\vec{y})\in L$ with $M \models \forall \vec{x},\vec{y},\vec{z} ( \psi(\vec{x},\vec{y}) \wedge \psi(\vec{x},\vec{z}) \rightarrow \vec{y} = \vec{z} )$. Every $M$-continuous sequence $\vec{a_0}, \vec{a_1}, \ldots, \vec{a_i}$ of vectors of numbers of $M$ with $M\models \forall j < i\ \psi(\vec{a_j}, \vec{a_{j+1}})$ can be coded as a number of $M$ by extended coding functions.
\end{M_continuous_computation_sequence_is_codified}
\begin{proof}
  Let $\varphi(\vec{x},y,\vec{z}) ::= \exists w ( \overline{(w)_0} = \vec{x} \wedge \forall u < y\ \psi(\overline{(w)_u}, \overline{(w)_{u+1}})  \wedge \overline{(w)_y} = \vec{z})$. Using Lemmas \ref{basic_properties_of_extended_pairing_functions} and \ref{basic_properties_of_extended_beta_function}, it's easy to see that $M \models \varphi(\vec{a_0},0,\vec{a_0}) \wedge \forall j < i ( \varphi(\vec{a_0},j,\vec{a_j}) \rightarrow \varphi(\vec{a_0},j+1,\vec{a_{j+1}}) )$. Since the sequence $\vec{a_0}, \vec{a_1}, \ldots, \vec{a_i}$ is $M$-continuous, we have that $M \models \varphi(\vec{a_0},0,\vec{a_0}) \wedge \forall j < i ( \varphi(\vec{a_0},j,\vec{a_j}) \rightarrow \varphi(\vec{a_0},j+1,\vec{a_{j+1}}) ) \rightarrow \forall j \leq i\ \varphi(\vec{a_0},j,\vec{a_j})$. Then it follows that $M \models \forall j \leq i\ \varphi(\vec{a_0},j,\vec{a_j})$. It's easy to prove, similar to the proof of Claim \ref{computable_sequence_generated_by_formulas} (1), that $M\models \forall j \leq i, \forall \vec{s},\vec{t} ( \varphi(\vec{a_0},j,\vec{s}) \wedge \varphi(\vec{a_0},j,\vec{t}) \rightarrow \vec{s} = \vec{t} )$. Then we have $M \models \forall j \leq i ( \exists ! \vec{z}\ \varphi(\vec{a_0},j,\vec{z}) \wedge  \varphi(\vec{a_0},j,\vec{a_j}) )$. By Theorem \ref{code_generated_by_formulas}, it follows that $\vec{a_0}, \vec{a_1}, \ldots, \vec{a_i}$ can be coded as a number of $M$ by extended coding functions.
\end{proof}

Note that, in Theorem \ref{M_continuous_computation_sequence_is_codified}, the premise $M \models \forall \vec{x},\vec{y},\vec{z} ( \psi(\vec{x},\vec{y}) \wedge \psi(\vec{x},\vec{z}) \rightarrow \vec{y} = \vec{z} )$ is essential for the conclusion to hold. Let $\psi(x,y) ::= y = 0 \vee y = 1$. Let $A$ be as defined in the proof of Proposition \ref{M_continuous_is_uncodified}. It's trivial that each element $<a_0,a_1,\ldots,a_i>$ of $A$ satisfies $M\models \forall j < i\ \psi(\vec{a_j}, \vec{a_{j+1}})$. However, one of elements of $A$ can't be coded as a number of $M$ by extended coding functions.

\newtheorem{equivalence_of_the_continuous_semantics_and_the_X_semantics_of_while_programs}[definition_of_continuous_sequences]{Theorem}
\begin{equivalence_of_the_continuous_semantics_and_the_X_semantics_of_while_programs}
  For every $S\in WP$ and every $M\models PA$, $A_S^M = C_S^M$.
\end{equivalence_of_the_continuous_semantics_and_the_X_semantics_of_while_programs}
\begin{proof}
Let $S\in WP$ and let $M\models PA$. The proof of $A_S^M = C_S^M$ proceeds by induction on $S$. The cases for assignment, composition and conditional are easy. For the iteration case, let $S\equiv while\ B\ do\ S_0\ od$ with $S_0\in WP$. To prove $A_S^M = C_S^M$, it suffices to prove that $A_S^M \subseteq C_S^M$ and $A_S^M \supseteq C_S^M$. By Claim \ref{computable_sequence_generated_by_formulas}, together with Proposition \ref{M_finiteness_is_M_continuous}, it's easy to see that $A_S^M \subseteq C_S^M$. To prove $A_S^M \supseteq C_S^M$, by Theorem \ref{arithmetical_definability_of_M_computable_functions}, it suffices to prove that $(v,v')\in C_S^M$ $\Rightarrow$ $M \models \alpha_S(v(\vec{x}), v'(\vec{x}))$. Suppose that $(v,v')\in C_S^M$. Then we have to prove that $M \models \alpha_S(v(\vec{x}), v'(\vec{x}))$. By the definition of $C_S^M$, we have that $\textbf{E}$ an $M$-continuous sequence $\vec{x_0}$, $\ldots$, $\vec{x_i} \in M$ $[$ $v(\vec{x}) = \vec{x_0}$ $\&$ $\textbf{A} j < i ( M,v(\vec{x_j}/\vec{x})\models B$ $\&$ $(v(\vec{x_j}/\vec{x}), v(\vec{x_{j+1}}/\vec{x}))\in C_{S_0}^M ) $ $\&$ $v' = v(\vec{x_i}/\vec{x})$ $\&$ $M,v'\not\models B$ $]$. By the induction hypothesis, together with Theorem \ref{arithmetical_definability_of_M_computable_functions}, it follows that $(v,v')\in C_{S_0}^M$ $\Leftrightarrow$ $M \models \alpha_{S_0}(v(\vec{x}), v'(\vec{x}))$. Then we have that $\textbf{E}$ an $M$-continuous sequence $\vec{x_0}$, $\ldots$, $\vec{x_i} \in M$ $[$ $v(\vec{x}) = \vec{x_0}$ $\&$ $\textbf{A} j < i ( M \models B(\vec{x_j})$ $\&$ $M \models \alpha_{S_0}(\vec{x_j}, \vec{x_{j+1}}) )$ $\&$ $v'(\vec{x}) = \vec{x_i}$ $\&$ $M\not\models B(v'(\vec{x}))$ $]$. By Theorem \ref{M_continuous_computation_sequence_is_codified}, together with the fact $M\models \forall \vec{x},\vec{y},\vec{z}(\alpha_{S_0}(\vec{x},\vec{y}) \wedge \alpha_{S_0}(\vec{x},\vec{z}) \rightarrow \vec{y} = \vec{z})$ (cf. Theorem \ref{arithmetical_definability_of_M_computable_functions}), we have that $\textbf{E}$ $i,w\in M$ $[$ $M\models v(\vec{x}) = \overline{(w)_0}$ $\&$ $M\models \forall j<i ( B(\overline{(w)_j}) \wedge \alpha_{S_0}(\overline{(w)_j}, \overline{(w)_{j+1}}) )$ $\&$ $M\models v'(\vec{x}) = \overline{(w)_i}$ $\&$ $M\not\models B(v'(\vec{x}))$ $]$. By the definition of $\alpha_S$, it follows that $M \models \alpha_S(v(\vec{x}), v'(\vec{x}))$.
\end{proof}

Summarizing the above results, we conclude that the X-semantics of while-programs is equivalent to the Hungary semantics. Moreover, the use of $M$-finiteness in the definition of the nonstandard semantics has an advantage over the use of $M$-continuousness: $M$-finiteness enables us to construct a $\Sigma_1$-formula to uniquely determine the sequence of computation, whereas in the Hungary semantics, we have to verify whether a sequence satisfies the induction axioms for all formulas.

\section{Soundness and completeness of Hoare logic}

\subsection{Soundness and Completenss of $HL(T)$}

\theoremstyle{definition}
\newtheorem{satisfiability_of_asserted_programs_in_models}{Definition}[subsection]
\begin{satisfiability_of_asserted_programs_in_models}\label{satisfiability_of_asserted_programs_in_models}
Given a model $M\models PA$ and an asserted program $\{p\}S\{q\}$, $\{p\}S\{q\}$ is satisfied at $M$ w.r.t. the X-semantics, denoted
\begin{equation*}
M\models_X \{p\}S\{q\},
\end{equation*}
iff for every assignment $v$ over $M$,
\begin{equation*}
M,v\models p \Rightarrow \textbf{A}v'\big((v,v')\in A_S^M \Rightarrow M,v'\models q\big).
\end{equation*}
\end{satisfiability_of_asserted_programs_in_models}

\newtheorem{satisfiability_of_asserted_programs_in_theories}[satisfiability_of_asserted_programs_in_models]{Definition}
\begin{satisfiability_of_asserted_programs_in_theories}\label{satisfiability_of_asserted_programs_in_theories}
Given a theory $PA\subseteq T\subseteq Th(N)$ and an asserted program $\{p\}S\{q\}$, $\{p\}S\{q\}$ is satisfied at $T$ w.r.t. the X-semantics, denoted
\begin{equation*}
HL(T) \models_X \{p\}S\{q\},
\end{equation*}
iff for every model $M\models T$,
\begin{equation*}
M \models_X \{p\}S\{q\}.
\end{equation*}
\end{satisfiability_of_asserted_programs_in_theories}

\newtheorem{definition_of_strongest_postconditions}[satisfiability_of_asserted_programs_in_models]{Definition}
\begin{definition_of_strongest_postconditions}\label{definition_of_strongest_postconditions}
Let $p\in L$, $S\in WP$ and $M\models PA$. The strongest postcondition $sp_X^M(p,S)$ of $S$ relative to $p$ w.r.t. the X-semantics on the set $V$ of assignments over $M$ is defined by
\begin{equation*}
 sp_X^M(p,S) ::= \{ v\in V : \textbf{E} v'\in V ( M,v'\models p ~\&~ (v',v)\in A_S^M ) \}.
\end{equation*}
The formal strongest postcondition $SP(p,S)(\vec{x})$ of $S$ relative to $p$ is defined by
\begin{equation*}
  SP(p,S)(\vec{x}) ::= \exists \vec{u} ( p(\vec{u}/\vec{x})\wedge \alpha_S(\vec{u}/\vec{x},\vec{x}/\vec{y}) ).
\end{equation*}
\end{definition_of_strongest_postconditions}

\theoremstyle{plain}
\newtheorem{strongest_postcondition}[satisfiability_of_asserted_programs_in_models]{Lemma}
\begin{strongest_postcondition}[Strongest Postcondition Lemma]\label{strongest_postcondition}
Let $p\in L$, $S\in WP$ and $M\models PA$. It is the case that

(i) $v\in sp_X^M(p,S)$ $\Leftrightarrow$ $M,v\models SP(p,S)$;

(ii) $M\models_X \{p\}S\{SP(p,S)\}$;

(iii) $M\models_X \{p\}S\{q\}$ $\Leftrightarrow$ $M\models SP(p,S)\rightarrow q$.
\end{strongest_postcondition}
\begin{proof}
  Immediate from Theorem \ref{arithmetical_definability_of_M_computable_functions}.
\end{proof}

\theoremstyle{definition}
\newtheorem{definition_of_loop_invariants}[satisfiability_of_asserted_programs_in_models]{Definition}
\begin{definition_of_loop_invariants}\label{definition_of_loop_invariants}
Let $p \in L$, $S ::= while\ B\ do\ S_0\ od$ and $M\models PA$. The loop invariant $inv_X^{M}(p,S)$ of $S$ relative to $p$ w.r.t. the X-semantics on the set $V$ of assignments over $M$ is defined by
\begin{align*}
  inv_X^{M}(p,S) ::= \{ & v\in V \ : \ \textbf{E}~i,\vec{x_0},\vec{x_1},\ldots,\vec{x_i}\in M ( M\models \forall j\leq i\ \varphi_{S_0}(\vec{x_0},j,\vec{x_j}) \\
                        & ~\&~ M,v(\vec{x_0}/\vec{x}) \models p ~\&~ \textbf{A}~ j < i ( M,v(\vec{x_j}/\vec{x})\models B \\
                        & ~\&~ (v(\vec{x_j}/\vec{x}), v(\vec{x_{j+1}}/\vec{x}) ) \in A_{S_0}^M ) ~\&~ v(\vec{x})= \vec{x_i} ) \}.
\end{align*}
The formal loop invariant $INV(p,S)(\vec{x})$ of $S$ relative to $p$ is defined by
\begin{equation*}
  INV(p,S)(\vec{x}) ::= \exists \vec{u} ( p(\vec{u}/\vec{x}) \wedge \exists i\ \alpha_S^*(i,\vec{u}/\vec{x},\vec{x}/\vec{y}) ).
\end{equation*}
\end{definition_of_loop_invariants}

\theoremstyle{plain}
\newtheorem{loop_invariant}[satisfiability_of_asserted_programs_in_models]{Lemma}
\begin{loop_invariant}[Loop Invariant Lemma]\label{loop_invariant}
Let $p \in L$, $S ::= while\ B\ do\ S_0\ od$ and $M\models PA$. It is the case that

(i) $v\in inv_X^{M}(p,S)$ $\Leftrightarrow$ $M,v\models INV(p,S)(\vec{x})$;

(ii) $M\models p\rightarrow INV(p,S)$;

(iii) $M \models_X \{INV(p,S) \wedge B\} S_0 \{INV(p,S)\}$;

(iv) $M\models_X \{p\}S\{q\}$ $\Leftrightarrow$ $M\models INV(p,S)(\vec{x}) \wedge \neg B(\vec{x}) \rightarrow q(\vec{x})$.
\end{loop_invariant}

\begin{proof}
  Follows from Theorem \ref{arithmetical_definability_of_M_computable_functions} and Corollary \ref{induction_of_loop_programs}.
\end{proof}

\theoremstyle{definition}
\newtheorem{soundness_and_completeness_of_Hoare_logic}[satisfiability_of_asserted_programs_in_models]{Definition}
\begin{soundness_and_completeness_of_Hoare_logic}
Given a theory $PA\subseteq T\subseteq Th(N)$, $HL(T)$ is sound and complete w.r.t. the X-semantics iff for any $p,q\in L$ and any $S\in WP$,
\begin{equation*}
HL(T)\vdash \{p\} S \{q\} \mbox{\ iff \ } HL(T)\models_X \{p\} S \{q\}.
\end{equation*}
\end{soundness_and_completeness_of_Hoare_logic}

We are now positioned to state the main theorem of this paper.

\theoremstyle{plain}
\newtheorem{soundness_and_completeness_theorem}[satisfiability_of_asserted_programs_in_models]{Theorem}
\begin{soundness_and_completeness_theorem}[Soundness and Completeness of $HL(T)$]\label{soundness_and_completeness_theorem}
Let $PA\subseteq T \subseteq Th(N)$. Then the Hoare logic $HL(T)$ is sound and complete w.r.t. the X-semantics. That is, for any $p,q\in L$ and any $S\in WP$,
\begin{equation*}
HL(T)\vdash \{p\} S \{q\} \mbox{\ iff \ } HL(T)\models_X \{p\} S \{q\}.
\end{equation*}
\end{soundness_and_completeness_theorem}
\begin{proof}
Let $p,q\in L$ and $S\in WP$. The proof of this theorem is partitioned into two parts: the ``only if'' part ($\Rightarrow$) and the ``if'' part ($\Leftarrow$).

$(\Rightarrow)$. Suppose that $HL(T)\vdash \{p\} S \{q\}$. Then we have to prove that $HL(T)\models_X \{p\} S \{q\}$. The argument proceeds by induction on the length of the proof of $\{p\} S \{q\}$ in $HL(T)$, where the last step is divided into the following cases.

Assignment Axiom Scheme: $S\equiv x_i := E$ and $HL(T)\vdash \{p\} S \{q\}$. By the Assignment Axiom Scheme, we have that $p \equiv q(E/x_i)$. This together with the X-semantics of $S$ (cf. Definition \ref{X_semantics_of_while_programs}) gives that $HL(T)\models_X \{p\} S \{q\}$.

Composition Rule: $S \equiv S_1;S_2$ and $HL(T) \vdash \{p\}S\{q\}$. By the Composition Rule, there exists $r(\vec{x})\in L$ such that $HL(T) \vdash \{p\} S_1 \{r\}$ and $HL(T) \vdash \{r\} S_2 \{q\}$. By the induction hypothesis, it follows that $HL(T) \models_X \{p\} S_1 \{r\}$ and $HL(T) \models_X \{r\} S_2 \{q\}$. By the X-semantics of $S$ (cf. Definition \ref{X_semantics_of_while_programs}), we have $HL(T)\models_X \{p\} S \{q\}$.

Conditional Rule: $S \equiv if\ B\ then\ S_1\ else\ S_2\ fi$ and $HL(T) \vdash \{p\} S \{q\}$. By the Conditional Rule, we have that $HL(T) \vdash \{p \wedge B\} S_1 \{q\}$ and $HL(T)\vdash  \{p \wedge \neg B \} S_2 \{q\}$. By the induction hypothesis, it follows that $HL(T) \models_X \{p \wedge B\} S_1 \{q\}$ and $HL(T)\models_X  \{p \wedge \neg B \} S_2 \{q\}$. By the X-semantics of $S$ (cf. Definition \ref{X_semantics_of_while_programs}), we have $HL(T)\models_X \{p\} S \{q\}$.

Iteration Rule: $S \equiv while\ B\ do\ S_0\ od$ and $HL(T) \vdash \{p\} S \{q\}$. By the Iteration Rule, we have that $HL(T) \vdash \{p \wedge B\} S_0 \{p\}$ and $q \equiv p \wedge \neg B$. By the induction hypothesis, it follows that $HL(T) \models_X \{p \wedge B\} S_0 \{p\}$. Let $M\models T$. By Theorem \ref{arithmetical_definability_of_M_computable_functions}, we have that $M \models p(\vec{x})\wedge B(\vec{x})\wedge \alpha_{S_0}(\vec{x},\vec{y})\rightarrow p(\vec{y}/\vec{x})$. Assuming that $M \models p(\vec{x}) \wedge \alpha_S(\vec{x},\vec{y})$, by the X-semantics of $S$ (cf. Definition \ref{X_semantics_of_while_programs}) and Theorem \ref{arithmetical_definability_of_M_computable_functions} there exist $i,$ $\vec{x_0}=\vec{x},$ $\vec{x_1},$ $\ldots,$ $\vec{x_i}=\vec{y}$ $\in$ $M$ with $M\models \forall j\leq i ( \exists! \vec{z}\ \varphi_{S_0}(\vec{x},j,\vec{z}) \wedge \varphi_{S_0}(\vec{x},j,\vec{x_j}) )$ such that $M\models \forall j<i ( B(\vec{x_j}) \wedge \alpha_{S_0}(\vec{x_j},\vec{x_{j+1}}) ) \wedge \neg B(\vec{x_i})$, then by induction on $j$ up to $i$ it's easy to see $M\models \forall j\leq i\ p(\vec{x_j}/\vec{x})$ and hence we have $M \models p(\vec{y}/\vec{x}) \wedge \neg B(\vec{y}/\vec{x})$, so $M\models q(\vec{y}/\vec{x})$. Then $M\models p(\vec{x}) \wedge \alpha_S(\vec{x},\vec{y})\rightarrow q(\vec{y}/\vec{x})$. By Theorem \ref{arithmetical_definability_of_M_computable_functions}, it follows that $M\models_X \{p\} S \{q\}$. Since $M\models T$ is arbitrary, we have that $HL(T)\models_X \{p\} S \{q\}$.

Consequence Rule: $HL(T) \vdash \{p\} S \{q\}$. By the Consequence Rule, there exist $p_1,q_1\in L$ such that $T \vdash p\rightarrow p_1$, $HL(T) \vdash \{p_1\}S\{q_1\}$ and $T \vdash q_1\rightarrow q$. By soundness of first-order logic, we have $T \models p\rightarrow p_1$ and $T \models q_1\rightarrow q$. By the induction hypothesis, it follows that $HL(T) \models_X \{p_1\}S\{q_1\}$. Then we have $HL(T)\models_X \{p\} S \{q\}$.

$(\Leftarrow)$. Suppose that $HL(T)\models_X \{p\} S \{q\}$. Then we have to prove that $HL(T)\vdash \{p\} S \{q\}$. The argument is by induction on the structure of $S$.

Assignment: $S \equiv x_i := E$. Suppose that $HL(T)\models_X \{p\} S \{q\}$. By the X-semantics of $S$ (cf. Definition \ref{X_semantics_of_while_programs}), we have that $T\models p\rightarrow q(E/x_i)$. By completeness of first-order logic, it follows that $T \vdash p\rightarrow q(E/x_i)$. By the Assignment Axiom Scheme, we have $HL(T) \vdash \{q(E/x_i)\} S \{q\}$. By the Consequence Rule, it follows that $HL(T) \vdash \{p\} S \{q\}$.

Composition: $S \equiv S_1;S_2$. Suppose that $HL(T)\models_X \{p\} S \{q\}$. By the Strongest Postcondition Lemma (cf. Lemma \ref{strongest_postcondition}), we have that $HL(T) \models_X \{p\} S_1 \{SP(p,S_1)\}$ and $HL(T) \models_X  \{SP(p,S_1)\} S_2 \{q\}$. Applying the induction hypothesis to $S_1$ and $S_2$ yields $HL(T) \vdash \{p\} S_1 \{SP(p,S_1)\}$ and $HL(T) \vdash  \{SP(p,S_1)\} S_2 \{q\}$. By the Composition Rule, it follows that $HL(T) \vdash \{p\} S \{q\}$.

Conditional: $S \equiv if\ B\ then\ S_1\ else\ S_2\ fi$. Suppose that $HL(T)\models_X \{p\} S \{q\}$. By the X-semantics of $S$ (cf. Definition \ref{X_semantics_of_while_programs}), we have that $HL(T) \models_X \{p \wedge B\} S_1 \{q\}$ and $HL(T) \models_X \{p \wedge \neg B\} S_2 \{q\}$. By the induction hypothesis applied to $S_1$ and $S_2$, it follows that $HL(T) \vdash \{p \wedge B\} S_1 \{q\}$ and $HL(T) \vdash \{p \wedge \neg B\} S_2 \{q\}$. By the Conditional Rule, we have $HL(T) \vdash \{p\} S \{q\}$.

Iteration: $S \equiv while\ B\ do\ S_0\ od$. Suppose that $HL(T)\models \{p\} S \{q\}$. By the Loop Invariant Lemma (cf. Lemma \ref{loop_invariant}), it follows that $T \models p \rightarrow INV(p,S)$, $HL(T) \models_X \{ INV(p,S) \wedge B \} S_0 \{ INV(p,S)\}$ and $T\models INV(p,S) \wedge \neg B \rightarrow q$. By completeness of first-order logic, we have that $T \vdash p \rightarrow INV(p,S)$ and $T \vdash INV(p,S) \wedge \neg B \rightarrow q$. Applying the induction hypothesis to $S_0$ yields $HL(T) \vdash \{ INV(p,S) \wedge B \} S_0 \{ INV(p,S)\}$. By the Iteration Rule, it follows that $HL(T) \vdash \{ INV(p,S) \} S \{ INV(p,S) \wedge \neg B \}$. By the Consequence Rule, we have that $HL(T) \vdash \{p\} S \{q\}$.

This completes the proof.
\end{proof}

\subsection{Reduction from $HL(T)$ to $T$}\label{section_of_from_hl_to_pa}

The reduction from $HL(T)$ to $T$ with $PA\subseteq T \subseteq Th(N)$ has been established by Bergstra and Tucker as the main theorem in \cite{bergstra_2}, and is shown as follows.

\newtheorem{bergstra_theorem}{Theorem}[subsection]
\begin{bergstra_theorem}\label{bergstra_theorem}
Given an assertion $p\in L$ and program $S\in WP$ one can effectively calculate
an assertion $SP(p,S)\in L$ such that

(1) $SP(p,S)$ defines the strongest postcondition of $S$ relative to $p$ on the set of states
over $N$;

(2) $HL(PA)\vdash \{p\}S\{SP(p,S)\}$;

\noindent And, for any refinement $T$ of Peano arithmetic, including $PA$ itself,

(3) $HL(T)\vdash \{p\}S\{q\}$ if and only if $T\vdash SP(p,S)\rightarrow q$.
\end{bergstra_theorem}

In Theorem \ref{bergstra_theorem}, statement (1) explains the meaning of $SP(p,S)$ in $N$; statement (2) is a preliminary result for statement (3), and it can be taken as a special case of (3) with $SP(p,S)$ in place of $q$; (3) is the core, and it establishes the reduction from $HL(T)$ to $T$ by using $SP(p,S)$.

The main results having been developed in this paper can be used to derive the above theorem:

\begin{proof}[Proof of Theorem \ref{bergstra_theorem}]
By Theorem \ref{arithmetical_definability_of_M_computable_functions} and Proposition \ref{comparison_of_two_semantics_over_N}, it follows that for any $\vec{n},\vec{m}\in N$, $f_S^N(\vec{n}) = \vec{m}$ iff $N\models \alpha_S(\vec{n},\vec{m})$. The formal strongest postcondition $SP(p,S)$, here, can be defined by $SP(p,S)(\vec{x}) ::= \exists \vec{u} ( p(\vec{u}/\vec{x}) \wedge \alpha_S(\vec{u}/\vec{x},\vec{x}/\vec{y}) )$. Then statement (1) of Theorem \ref{bergstra_theorem} follows. To prove Theorem \ref{bergstra_theorem}, it remains to prove statement (3). Let $PA\subseteq T \subseteq Th(N)$, $p,q\in L$, and $S\in WP$. Consider $HL(T)\vdash \{p\}S\{q\}$ as follows: by Theorem \ref{soundness_and_completeness_theorem}, it is equivalent to $HL(T)\models_X \{p\}S\{q\}$; by Theorem \ref{arithmetical_definability_of_M_computable_functions}, it is equivalent to $T\models p(\vec{x})\wedge \alpha_S(\vec{x},\vec{y})\rightarrow q(\vec{y}/\vec{x})$; by definition of $SP(p,S)$, it is equivalent to $T\models SP(p,S)\rightarrow q$; by soundness and completeness of first-order logic, it is equivalent to $T\vdash SP(p,S)\rightarrow q$.
\end{proof}

Observing the axiom system of $HL(T)$, we remark that the crucial and perhaps the most difficult step of finding a proof of $\{p\}S\{q\}$ in $HL(T)$ is to guess the corresponding intermediate assertions when applying the Composition and Iteration Rules. As Lemmas \ref{strongest_postcondition}, \ref{loop_invariant} and the proof of Theorem \ref{soundness_and_completeness_theorem} show, the intermediate assertions involved in this proof can be easily constructed, and their roles in guaranteeing the correctness of Theorem \ref{bergstra_theorem} are implicit. The real strength of Theorem \ref{bergstra_theorem} is due to the fact that it provides us a readily-operating means to deciding whether $\{p\}S\{q\}$ is provable in $HL(T)$: one simply needs to decide whether $SP(p,S)\rightarrow q$ is provable in $T$.

\subsection{Comparison with Floyd logic}\label{comparison_with_floyd_hoare_logic}

This Subsection is considered as a continuation to Subsection \ref{comparison_with_the_Hungary_semantics}. The following definitions concerning Floyd logic come from \cite[p165-166]{andreka_3}.

Let $PA\subseteq T \subseteq Th(N)$, $M\models T$, $P ::= <(i_0:u_0),...,(i_n:u_n)>$ and $\psi\in L$.

The dynamic formula $\Box(P,\psi)$ is said to be valid in $M$ w.r.t. continuous traces if ($\ast$) below holds.
\begin{quote}
($\ast$) For any continuous trace $<\vec{s_a}>_{a\in M}$ of $P$ in $M$ and for any $a\in M, \vec{s_a}(\lambda)\not\in\{i_m: m\leq n\}$
implies $M\models\psi[\vec{s_a}].$
\end{quote}
$M\models \Box(P,\psi)$ denotes that $\Box(P,\psi)$ is valid in $M$ w.r.t continuous traces; $T\models \Box(P,\psi)$ denotes that for every $M\models T$, $M\models \Box(P,\psi)$.

The Floyd logic serves to derive statements $\Box (P, \psi)$ from $T$. The set $lab(P)$ of labels of $P$ is defined as follows:
\begin{equation*}
  lab(P) ::= \{i_{m} : m\leq n+1\} \cup \{v : \exists m\leq n ( u_m = if\ E\ goto\ v )\}.
\end{equation*}
A Floyd derivation of $\Box(P,\psi)$ from $T$ consists of: a mapping $\Phi: lab(P) \rightarrow L$ together with the classical first-order derivations listed in (i)-(iv) below. Notation: When $z \in lab(P)$ we write $\Phi_z$ instead of $\Phi(z)$.

\begin{quote}
(i) A derivation: $T\vdash\Phi_{i_0}$.

(ii) To each command $i_m: x_j\hookleftarrow E$ occurring in $P$, a derivation:
$T\vdash \Phi_{i_m}\rightarrow \Phi_{i_{m+1}}(E/x_j)$.

(iii) To each command $i_m: if\ B\ goto\ v$ occurring in $P$, two derivations:
$T\vdash (B\land\Phi_{i_m})\rightarrow\Phi_v$ and $T\vdash
(\lnot B\land\Phi_{i_m})\rightarrow\Phi_{i_{m+1}}$.

(iv) To each $z\in (lab(P)- \{i_m : m\leq n\})$, a derivation:
$T\vdash \Phi_z\rightarrow\psi$.
\end{quote}
Now the existence of a Floyd derivation of $\Box(P,\psi)$ from
$T$ is denoted by $T\vdash\Box(P, \psi)$. The soundness and completeness of Floyd logic relative to $T$ has been established by H.~Andr\'{e}ka, et.al., as the main theorem in \cite{andreka_3}, and is shown as follows.

\theoremstyle{plain}
\newtheorem{soundness_and_completeness_of_Floyd_Hoare_logic}{Theorem}[subsection]
\begin{soundness_and_completeness_of_Floyd_Hoare_logic}[Soundness and completeness of Floyd logic]\label{soundness_and_completeness_of_Floyd_Hoare_logic}
  Let $PA \subseteq T \subseteq Th(N)$. For every labeled program $P$, and every $\psi \in L$, it is the case that $T\vdash\Box(P, \psi)$ iff $T\models\Box(P, \psi)$.
\end{soundness_and_completeness_of_Floyd_Hoare_logic}

To build the relationship of Hoare logic with Floyd logic, we remark that for each $S\in WP$, there is a labeled program $P$ such that the X-semantics of $S$ is equivalent to the semantics of $P$ in $PA$, and vice versa.

\newtheorem{relationship_of_Hoare_logic_with _Floyd_Hoare_logic}[soundness_and_completeness_of_Floyd_Hoare_logic]{Theorem}
\begin{relationship_of_Hoare_logic_with _Floyd_Hoare_logic}\label{relationship_of_Hoare_logic_with _Floyd_Hoare_logic}
  Let $PA \subseteq T \subseteq Th(N)$. For every $S\in WP$, every labeled program $P$, and every $q\in L$, if the X-semantics of $S$ is equivalent to the semantics of $P$ in $PA$, then it is the case that $HL(T)\vdash \{true\} S \{q\}$ iff $T\vdash \Box(P,q)$.
\end{relationship_of_Hoare_logic_with _Floyd_Hoare_logic}
\begin{proof}
  Immediate from Theorems \ref{soundness_and_completeness_theorem} and \ref{soundness_and_completeness_of_Floyd_Hoare_logic}.
\end{proof}

Intuitively Hoare logic can be thought of as a special first-order multi-modal logic with programs viewed as modalities; and the axiom system of Floyd logic looks more like a first-order logical system. Although these two axiom systems appear different, they are essentially the same first-order logical system, as Theorems \ref{relationship_of_Hoare_logic_with _Floyd_Hoare_logic} and \ref{bergstra_theorem} show.

\section{The X-semantics revisited}

\subsection{Comparison with the axiomatic semantics}\label{comparison_with_the_axiomatic_semantics}

The axiomatic semantics of while-programs is a relational semantics determined by Hoare logic. The following definition is as in \cite[p303]{bergstra_2} except that the side effects on non-program variables are explicitly ruled out.

\theoremstyle{definition}
\newtheorem{definition_of_axiomatic_semantics}{Definition}[subsection]
\begin{definition_of_axiomatic_semantics}\label{definition_of_axiomatic_semantics}
  Let $L'$ be a first-order logical language, let $WP'$ be the set of while-programs based on $L'$, and let $T$ be a first-order theory of $L'$. For any $S\in WP'$ with program variables $\vec{x}$, the axiomatic semantics $AX(T)_M(S)$ of $S$ in a model $M$ of $T$ is a binary relation on the set $V$ of assignments over $M$ defined as follows:
  \begin{eqnarray*}
   AX(T)_M(S) &::=& \{ (\sigma, \tau)\in V\times V: \textbf{A} y\not\in \vec{x}( \sigma(y) = \tau(y) ), \\
              && \textbf{A} p,q\in L', v\in V ( HL(T)\vdash \{p\}S\{q\}\\
   && \& \  M,v(\sigma(\vec{x})/\vec{x}) \models p \Rightarrow M,v(\tau(\vec{x})/\vec{x}) \models q ) \}.
  \end{eqnarray*}
\end{definition_of_axiomatic_semantics}

Observe from the above definition that $AX(T)_M(S)$ is the maximal input-output relation for $S$ over $M$ such that $HL(T)$ is sound in $M$ with respect to all the specifications of $S$. Although this definition of axiomatic semantics appears mathematically elegant, it is too abstract, and the general properties of its are not always nice, e.g. nondeterminism. (For more details, the reader refers to \cite{bergstra_2}.) Surprisingly, when restricting the first-order theory $T$ to $PA \subseteq T \subseteq Th(N)$, the axiomatic semantics is exactly what we pursue in this paper.

\theoremstyle{plain}
\newtheorem{equivalence_with_axiomatic_semantics}[definition_of_axiomatic_semantics]{Theorem}
\begin{equivalence_with_axiomatic_semantics}
  Let $PA \subseteq T \subseteq Th(N)$. For every $M\models T$, and every $S\in WP$, $AX(T)_M(S) = A_S^M$.
\end{equivalence_with_axiomatic_semantics}
\begin{proof}
  Let $M\models T$, let $V$ be the set of assignments over $M$, and let $S\in WP$ with program variables $\vec{x}$. To prove $AX(T)_M(S) = A_S^M$, it suffices to prove that $AX(T)_M(S) \supseteq A_S^M$ and $AX(T)_M(S) \subseteq A_S^M$.

  $(\supseteq)$. Let $(\sigma, \tau)\in A_S^M$. By Definition \ref{X_semantics_of_while_programs}, it's easy to see that $\textbf{A} y\not\in \vec{x} ( \sigma(y) = \tau(y) )$. To prove $(\sigma, \tau)\in AX(T)_M(S)$, by Definition \ref{definition_of_axiomatic_semantics}, it suffices to prove that $\textbf{A} p,q\in L, v\in V ( HL(T)\vdash \{p\}S\{q\}\ \& \  M,v(\sigma(\vec{x})/\vec{x}) \models p \Rightarrow M,v(\tau(\vec{x})/\vec{x}) \models q )$. Let $p,q\in L$ and $v\in V$ with $HL(T)\vdash \{p\}S\{q\}$ and $M,v(\sigma(\vec{x})/\vec{x}) \models p$. It remains to show that $M,v(\tau(\vec{x})/\vec{x}) \models q$. By soundness of $HL(T)$ w.r.t. the X-semantics (cf. Theorem \ref{soundness_and_completeness_theorem}), it follows that $HL(T) \models_X \{p\}S\{q\}$. Then we have $M \models_X \{p\}S\{q\}$. From the premise $(\sigma, \tau)\in A_S^M$, by Definition \ref{X_semantics_of_while_programs}, it follows that $(v(\sigma(\vec{x})/\vec{x}), v(\tau(\vec{x})/\vec{x}))\in A_S^M$. Since $M,v(\sigma(\vec{x})/\vec{x}) \models p$, we have that $M,v(\tau(\vec{x})/\vec{x}) \models q$.

  $(\subseteq)$. Let $(\sigma, \tau)\in AX(T)_M(S)$. Let $p(\vec{u},\vec{x}) ::= \vec{u} = \vec{x}$, $q(\vec{u},\vec{x}) ::= \alpha_S(\vec{u},\vec{x})$, and $v ::= \sigma(\sigma(\vec{x})/\vec{u})$, where $\vec{u}$ is disjoint from $\vec{x}$. Then we have $T \models p(\vec{u},\vec{x}) \wedge \alpha_S(\vec{x},\vec{y})\rightarrow q(\vec{u},\vec{y})$. By Theorem \ref{arithmetical_definability_of_M_computable_functions}, it follows that $HL(T)\models_X \{p\}S\{q\}$. Using completeness of $HL(T)$ w.r.t. the X-semantics (cf. Theorem \ref{soundness_and_completeness_theorem}), we have that $HL(T)\vdash \{p\}S\{q\}$. By definitions of $v$ and $p$, we have that $M,v(\sigma(\vec{x})/\vec{x}) \models p$. By the definition of $AX(T)_M(S)$ (cf. Definition \ref{definition_of_axiomatic_semantics}), it follows that $\textbf{A} y\not\in \vec{x} ( \sigma(y) = \tau(y) )$ and $M,v(\tau(\vec{x})/\vec{x}) \models q$. By definitions of $v$ and $q$, we have $M\models \alpha_S(\sigma(\vec{x}),\tau(\vec{x}))$. By Theorem \ref{arithmetical_definability_of_M_computable_functions}, it follows that $(\sigma,\tau)\in A_S^M$.
\end{proof}

Observe from the foregoing theorem that the axiomatic semantics of while-programs on every model of $T$ with $PA\subseteq T \subseteq Th(N)$ is independent of the choice of $T$. It has been justified by the Uniqueness Theorem \cite[p304]{bergstra_2} that in a general setting, the axiomatic semantics of while-programs on a structure $M$ of a first-order theory $T$ is independent of the choice of $T$. We can now say that for any $M\models PA$, the axiomatic semantics of while-programs on $M$ coincides with the X-semantics of while-programs on $M$.

\subsection{Usefulness of nonstandard semantics in Hoare logic}

Properties of nonstandard models are sometimes very useful to decide whether a Hoare's triple is provable, especially when such a triple is true in the standard model. Let $PA \subseteq T \subseteq Th(N)$, $p,q \in L$, and $S\in WP$. Suppose that $N \models_S \{p\}S\{q\}$ (or equivalently $N \models_X \{p\}S\{q\}$) and $HL(T)\not\vdash \{p\}S\{q\}$. Then, by Theorem \ref{soundness_and_completeness_theorem}, there should be a nonstandard $M\models T$ such that $M \not\models_X \{p\}S\{q\}$. This means that the X-semantics of while-programs in $M$ could be useful in deciding whether $\{p\}S\{q\}$ is provable in $HL(T)$. To show the usefulness of nonstandard semantics in Hoare logic, the following example is illustrated.

\theoremstyle{plain}
\newtheorem{incompleteness_of_hl}{Theorem}[subsection]
\begin{incompleteness_of_hl}[Cf. {\cite[Theorem 4.3]{bergstra_1}}]\label{incompleteness_of_hl}
Let $S ::= y:=0; while\ y<x\ do\ y:=y+1\ od$, let $T \supseteq PA$ such that $Thm(T)\subsetneqq Th(N)$, and let $\varphi(x)\in L$ such that $T\vdash \varphi(n)$ for each $n\in N$ and $T\nvdash \forall x\ \varphi(x)$ (There always exists such a $\varphi$ for any such $T$). It is the case that $HL(T)\models_S \{\neg\varphi(x)\} S \{false\}$ but $HL(T)\nvdash \{\neg\varphi(x)\} S \{false\}$.
\end{incompleteness_of_hl}

Theorem \ref{incompleteness_of_hl} shows the incompleteness of $HL(T)$ with $T \supseteq PA$ and $Thm(T)\subsetneqq Th(N)$ w.r.t. the standard semantics. The original proof of Theorem \ref{incompleteness_of_hl} in \cite[the proof of Theorem 4.3]{bergstra_1} exploits the proof concept of $HL(T)$ to derive a contradiction, yet it seems to be cumbersome. Using the X-semantics of while-programs, a concise proof of Theorem \ref{incompleteness_of_hl} is shown below.

\begin{proof}[Proof of Theorem \ref{incompleteness_of_hl}]
  Let $S$, $T$ and $\varphi(x)$ be as defined in Theorem \ref{incompleteness_of_hl}. It's trivial that $HL(T)\models_S \{\neg\varphi(x)\} S \{false\}$. Thus it remains to prove $HL(T)\nvdash \{\neg\varphi(x)\} S \{false\}$. Since $N \models_X \{\neg\varphi(x)\} S \{false\}$, by Theorem \ref{soundness_and_completeness_theorem}, it suffices to prove that for some nonstandard $M\models T$, $M \not\models_X \{\neg\varphi(x)\} S \{false\}$. By the definition of $\varphi(x)$, together with completeness of first-order logic, it follows that there exists a nonstandard $M\models T$ such that $M\models \exists x\ \neg \varphi(x)$. Let $M$ be such a model. By Example \ref{example_for_two_semantics}, we have that for any $x,y\in M$, $g_S^M(x,y) = (x,x)$. Then $M \not\models_X \{\neg\varphi(x)\} S \{false\}$ follows.
\end{proof}

\section{Remarks on the completeness issues}

The most popular completeness form we have known for Hoare logic is the relative completeness (for a complete knowledge of relative completeness, the reader refers to the distinguished survey article \cite{apt_1}):
\begin{quote}
  Let $L'$ be a first-order logical language, and let $M$ be a model of $L'$. Hoare logic is complete for while-programs relative to $M$ iff for all asserted programs $\Phi$, if $M\models_H \Phi$, then $Th(M)\vdash_H \Phi$, where $\models_H$ and $\vdash_H$ denote the satisfaction and derivation relations of Hoare logic.
\end{quote}

Cook \cite{cook_1} considered the relative completeness of Hoare logic with the expressiveness condition (for the incompleteness due to inexpressiveness, and the completeness with inexpressiveness, the reader refers to \cite{wand_1} and \cite{bergstra_3} respectively):
\begin{quote}
  Let $L'$ and $M$ be as defined above. If $L'$ is expressive for while-programs relative to $M$ (i.e., for all assertions $p\in L'$ and all while-programs $W$ based on $L'$, there exists an assertion $q\in L'$ that defines the strongest postcondition of $W$ relative to $p$ on the set of assignments over $M$), then Hoare logic is complete for while-programs relative to $M$.
\end{quote}

Unfortunately, there are only two kinds of expressive structures for while-programs w.r.t. the standard semantics: the standard model $N$ and finite structures \cite{lipton_1}. In contrast, as Theorem \ref{arithmetical_definability_of_M_computable_functions} and Lemmas \ref{strongest_postcondition}, \ref{loop_invariant} imply, $L$ is expressive for while-programs w.r.t. the X-semantics relative to every model of $PA$, so the new semantics of Hoare logic extends the applicability of Cook's completeness theorem from $N$ to nonstandard models of $PA$. Except for the paucity of expressive structures, the relative completeness of Hoare logic has a theoretical weakness that we can't come to terms with: for each expressive structure $M$ it has to pick out the complete theory $Th(M)$ as the specification axioms to derive the whole set of true asserted programs in $M$. Consider for example the case $M\equiv N$. Let $T \supseteq PA$ with $Thm(T)\subsetneqq Th(N)$. Then there exists a sentence $p\in L$ such that $N\models p$ and $T\nvdash p$. It's easy to see that $N\models_S \{true\} x := x \{p\}$ (or equivalently $N\models_X \{true\} x := x \{p\}$) and $HL(T)\nvdash \{true\} x := x \{p\}$ (by the choice of $p$ and completeness of first-order logic, there exists $M\models T$ such that $M\nvDash p$; then we have $HL(T)\nvDash_X \{true\} x := x \{p\}$; by completeness of $HL(T)$ w.r.t. the X-semantics, it follows that $HL(T)\nvdash \{true\} x := x \{p\}$). From the above analysis, we achieve that $Th(N)$ is the only extension $T$ of $PA$ such that $HL(T)$ is complete relative to $N$ w.r.t. both the standard semantics and the X-semantics. However, by G\"{o}del's incompleteness theorem, $Th(N)$ is not recursively enumerable (even not arithmetical \cite[Lemma 17.3]{c. and l.}), which means that it is far beyond the power of any mechanical device to enumerate the elements of $Th(N)$. To fix up this drawback, we have studied the completeness issues of Hoare logic relative to $N$ by restricting assertions to subclasses of arithmetical formulas \cite{xu17} (and by restricting inputs to $N$ \cite{xu_3}): arithmetical extensions of $PA$ suffice to act as the assertion theory, and the lower the level of the assertions in the arithmetical hierarchy the lower the level of the required assertion theory is.

Historically, logical completeness of the first-order Hoare logic, which as an important completeness form stems from completeness of first-order logic, was studied extensively:
\begin{quote}
  Let $L'$ be a first-order logical language, and let $T$ be a first-order theory of $L'$. Hoare logic is complete for while-programs relative to $T$ iff for all asserted programs $\Phi$, if $T\models_H \Phi$, then $T\vdash_H \Phi$, where $T\models_H \Phi$ denotes that for any $M\models T$, $M\models_H \Phi$.
\end{quote}

Bergstra and Tucker \cite{bergstra_3} have pointed out that $Th(N)$ is the unique extension $T$ of $PA$ for which $HL(T)$ is logically complete w.r.t. the standard semantics. (The incompleteness of $HL(T)$ with $T \supseteq PA$ and $Thm(T)\subsetneqq Th(N)$ w.r.t. the standard semantics has been shown in Theorem \ref{incompleteness_of_hl}.) To remedy this unsatisfying completeness result, we have investigated the logical completeness of Hoare logic with inputs over the standard model \cite{xu16}: a particular extension $PA^+$ of $PA$ suffices to prove all valid Hoare's triples requiring that the input has a standard value; moreover $PA^+$ is shown to be minimal with the property. To see the essential difference of the completeness theorem of Bergstra and Tucker from that presented in this paper, we remark that the only factor leading to the difference is the semantics. Let $T \supseteq PA$ with $Thm(T)\subsetneqq Th(N)$. By Propositions \ref{comparison_of_two_semantics_over_N} and \ref{comparison_of_two_semantics_over_M}, for any $M\models T$, and any $S\in WP$, we have that $R_S^M$ is a (sometimes proper) subset of $A_S^M$. By the semantics of Hoare logic, it follows that $\{ \{p\}S\{q\} : p,q\in L, S\in WP, HL(T) \models_X \{p\}S\{q\} \}$ is properly contained in $\{ \{p\}S\{q\} : p,q\in L, S\in WP, HL(T) \models_S \{p\}S\{q\} \}$. As is shown in Subsection \ref{comparison_with_the_axiomatic_semantics}, the X-semantics of while-programs is an axiomatic semantics, which means that for any $M\models PA$ and any $S\in WP$, $A_S^M$ is the maximal input-output relation for $S$ over $M$ such that $HL(T)$ is sound in $M$ with respect to all the specifications of $S$ (in contrast, $R_S^M$ is not always maximal in this sense). Hence among all the semantics of while-programs that make $HL(T)$ sound in every model of $T$ (including the standard semantics), the axiomatic semantics makes the set $\{ \{p\}S\{q\} : p,q\in L, S\in WP, HL(T) \models_X \{p\}S\{q\} \}$ be minimal and naturally coincide with the set $\{ \{p\}S\{q\} : p,q\in L, S\in WP, HL(T) \vdash \{p\}S\{q\} \}$. Furthermore, Bergstra and Tucker established the logical completeness of Hoare logic based on the axiomatic semantics in a general setting \cite{bergstra_1}.

The Hungary School \cite{andreka_1,andreka_2,andreka_3} provides an alternative approach to characterizing the maximal semantics of programs for which Floyd logic is sound, i.e. the Hungary semantics, and shows that Floyd logic is logically complete w.r.t. this semantics. In Subsections \ref{comparison_with_the_Hungary_semantics} and \ref{comparison_with_floyd_hoare_logic}, a detailed comparison of this technique with that presented in this paper has been given. Csirmaz \cite{csirmaz_1}, and Hortal\'{a}-Gonz\'{a}lez et.al. \cite{hortala_1} extended this technique into a general situation. For the relationship of the general Floyd logic with weak second order logic, the reader refers to the literature \cite{makowsky_1}. Moreover, Andr\'{e}ka et.al. \cite{andreka_4}, and Pasztor \cite{pasztor_1} used many-sorted logic (i.e. Henkin semantics \cite{henkin_1}) to define Floyd-Hoare logic, thus obtaining the completeness of Floyd-Hoare logic immediately from that of many-sorted logic.

Kozen and Tiuryn \cite{kozen_1} viewed Hoare logic as a special propositional modal logic by abstracting assertions and programs to propositional symbols, and gained the completeness of this propositional Hoare logic.

\section{Conclusion}
In this paper, for nonstandard models $M$ of $PA$, we have introduced the concept of $M$-finiteness as the desired finiteness as far as $M$ is concerned; we have exhibited an explicit axiomatic semantics for while-programs in nonstandard models of $PA$, and proven that it coincides with the Hungary semantics; based on this semantics, we have defined a new semantics of Hoare logic, and shown the soundness and completeness of its axiom system w.r.t. the new semantics; various comparisons have shown that our approach is not only mathematically elegant, but also practically useful.

\section*{Acknowledgement}
The authors would like to thank the 973 Program of China (Grant No. 2014CB340701), the National Natural Science Foundation of China (Grant Nos. 61672504 and 61472474), and the CAS-SAFEA International Partnership Program for Creative Research Teams for the financial support. The authors would also like to thank the anonymous referees for their valuable suggestions and comments that helped improving this paper.

\end{document}